\documentclass[sigconf]{acmart}
\usepackage{multirow}
\usepackage{algorithm}
\usepackage{algpseudocode}
\usepackage{xspace}
\usepackage{mathtools, nccmath}

\DeclarePairedDelimiter{\nint}\lfloor\rceil
\algrenewcommand\algorithmicrequire{\textbf{Input:}}
\algrenewcommand\algorithmicensure{\textbf{Output:}}
\AtBeginDocument{%
  \providecommand\BibTeX{{%
    \normalfont B\kern-0.5em{\scshape i\kern-0.25em b}\kern-0.8em\TeX}}}

\newcommand{\minisection}[1]{\vspace{0.05in}\noindent {\bf #1}}
\newcommand{\ourmethod}{ALISE\xspace}
\setcopyright{acmlicensed}
\copyrightyear{2024}
\acmYear{2024}
\acmDOI{10.1145/3676536.3676659}

\acmConference[ICCAD '24]{
the 43rd IEEE/ACM International Conference on Computer-Aided Design
}{Oct 27--31,
  2024}{New Jersey, USA}
\acmISBN{978-1-4503-XXXX-X/24/10}

\begin{document}

%%
%% The "title" command has an optional parameter,
%% allowing the author to define a "short title" to be used in page headers.
\title{ALISE: Accelerating Large Language Model Serving with Speculative Scheduling}

% \title{MeRino: Automatic Design for Generative \\ Language Models on IoT Devices}

% \title{Enabling Efficient Large Language Models (LLMs) on Edge Devices: A Zero-Shot NAS Approach}

%%
%% The "author" command and its associated commands are used to define
%% the authors and their affiliations.
%% Of note is the shared affiliation of the first two authors, and the
%% "authornote" and "authornotemark" commands
%% used to denote shared contribution to the research.

\author{Youpeng Zhao}
\affiliation{%
  \institution{University of Central Florida}
  \city{Orlando, FL}
  \country{USA}}
\email{youpeng.zhao@ucf.edu}

\author{Jun Wang}
\affiliation{%
  \institution{University of Central Florida}
  \city{Orlando, FL}
  \country{USA}}
\email{jun.wang@ucf.edu}

%%
%% By default, the full list of authors will be used in the page
%% headers. Often, this list is too long, and will overlap
%% other information printed in the page headers. This command allows
%% the author to define a more concise list
%% of authors' names for this purpose.
% \renewcommand{\shortauthors}{Trovato and Tobin, et al.}

%%
%% The abstract is a short summary of the work to be presented in the
%% article.
\begin{abstract}
Large Language Models (LLMs) represent a revolutionary advancement in the contemporary landscape of artificial general intelligence (AGI). 
As exemplified by ChatGPT, LLM-based applications necessitate minimal response latency and maximal throughput for inference serving. 
However, due to the unpredictability of LLM execution, the first-come-first-serve (FCFS) scheduling policy employed by current LLM serving systems suffers from head-of-line (HoL) blocking issues and long job response times. 

In this paper, we propose a new efficient LLM inference serving framework, named ALISE.
The key design paradigm of ALISE is to leverage a novel speculative scheduler by estimating the execution time for each job and exploiting such prior knowledge to assign appropriate job priority orders, thus minimizing potential queuing delays for heterogeneous workloads.
Furthermore, to mitigate the memory overhead of the intermediate key-value (KV) cache, we employ a priority-based adaptive memory management protocol and quantization-based compression techniques.
Evaluations demonstrate that in comparison to the state-of-the-art solution vLLM, ALISE improves the throughput of inference serving by up to $1.8\times$ and $2.1\times$ under the same latency constraint on the Alpaca and ShareGPT datasets, respectively.
\end{abstract}

\maketitle

\section{Introduction}
In recent years, large language models (LLMs) have demonstrated ubiquitous performance across various natural language processing (NLP) tasks~\cite{gpt3,llama2, opt,llama3}. 
Notably, LLMs can generalize well to perform multiple tasks without additional fine-tuning, which makes them versatile and adaptable for downstream AI applications~\cite{chatgpt}. 
Inference serving is crucial to applications driven by LLMs, such as AI chatbots. 
The interactive nature of such applications requires the system to produce inference results for user requests at low latency and high throughput.

The inference process of LLMs exhibits the unique characteristics of \textit{autoregressiveness}, where LLMs generate new tokens sequentially based on the input (prompt) tokens and all prior generated tokens. 
LLM inference typically requires running multiple iterations of the model to complete the token generation.
General DNN inference serving systems~\cite{clockwork,shepherd,clipper,infaas} focus on deterministic workloads such as image classification and object detection, which are highly predictable. 
The autoregressive nature makes these solutions not work and scale well for LLM workloads, as in LLM inference, each token is generated based on the preceding ones, and this sequential dependency can result in unpredictable execution runtime for generated sequences. 

Existing LLM systems focus on improving serving throughput performance with custom CUDA kernels~\cite{fastertransformer,flashattention,flashattention2}, iteration-level scheduling~\cite{orca,fastserve}, and efficient memory management~\cite{vLLM,flexgen,alisa}.
Unfortunately, despite large strides toward improving the performance of serving LLMs, today's system platforms continue to struggle to provide low latency, especially in real-world workloads.
We argue that this struggle stems from the non-preemptive first-come-first-served (FCFS)~\cite{fcfs} scheduling strategy, which causes head-of-line (HoL) blocking issues when processing heterogeneous workloads.
Run-to-completion scheduling is recognized for its HoL blocking issue, which is especially troublesome for generative inference tasks. 
This occurs when prior scheduled tasks, potentially lengthy in execution, hinder the timely processing of subsequent shorter tasks. 
Such queuing delays could significantly deteriorate the quality of service (QoS) provided. 
How to efficiently handle heterogeneous requests with variable execution time and perform appropriate scheduling to maintain high throughput remains a challenging problem.

\begin{figure*}[!t]
\begin{center}
  \includegraphics[width=\linewidth]{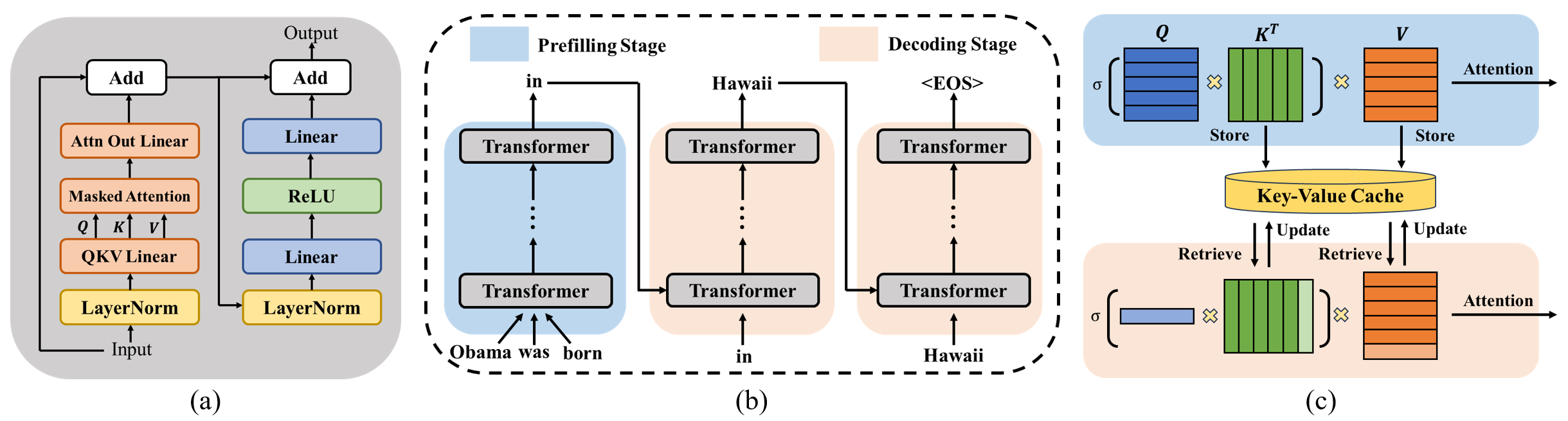}
\end{center}
\vspace{-4mm}
  \caption{(a) Operations in the transformer layer. 
  (b) An illustrative example of the autoregressive LLM inference process.
  (c) KV cache mechanism: at the prefilling stage, all input tokens are processed simultaneously, and the KV cache is initialized; 
  at the decoding stage, the stored KV cache is retrieved for reuse and updated by iteration until termination.
  }
\label{fig:background}
\end{figure*}

In this paper, we introduce \ourmethod, to \underline{a}ccelerate \underline{L}LM serv\underline{i}ng with a novel \underline{s}peculative sch\underline{e}duler.
A key insight behind \ourmethod is to estimate the execution time of heterogeneous LLM requests.
Specifically, we adopt a retrieval-based length predictor, where a user prompt is first embedded into high-dimension vectors using a pre-trained text encoder, and the output length is accurately predicted using an ensemble of a query database and an all-MLP decoder.
Based on such prior knowledge, we can estimate each job's execution time and leverage priority queues to perform preemptive scheduling, thus alleviating HoL issues and minimizing the total response time. 
To efficiently manage the intermediate KV cache of preempted jobs, we design a priority-based adaptive memory management mechanism that dynamically performs swapping operations (e.g., upload and offload) of the KV cache for preempted jobs based on each job's estimated wait time (EWT). 
Furthermore, we utilize the quantization technique to compress the KV cache to lower precision further reducing the overall memory overhead.

We implement and evaluate \ourmethod across different serving scenarios.
Extensive evaluations demonstrate that \ourmethod outperforms the existing FCFS solutions in end-to-end latency and overall throughput. 
Specifically, \ourmethod sustains up to $1.8\times$ and $2.1\times$ throughput improvement against the state-of-the-art solution, vLLM~\cite{vLLM}, on the Alpaca~\cite{alpaca} and ShareGPT~\cite{shareGPT} datasets, respectively.

In summary, we make the following contributions:
\begin{itemize}
    \item We propose a novel retrieval-based speculative scheduler to estimate the execution time of each incoming job and schedule workloads based on their priorities to minimize the response latency.
    \item We design an adaptive memory manager 
    that intelligently performs preemptive offload/upload operations for unused intermediate KV cache and quantization to reduce memory overhead.
    \item We evaluate our serving solution, termed \ourmethod, on different scenarios and show that it significantly outperforms the existing LLM inference solutions, such as ORCA~\cite{orca} and vLLM~\cite{vLLM}.
\end{itemize}
The remainder of this paper is organized as follows. Section~\ref{sec:background_motivation} presents the background and motivation for our work.
Then, Section~\ref{sec:design} elaborates on the methodology of our system design, with evaluation followed in Section~\ref{sec:evaluation}.
Finally, Section~\ref{sec:related_work} recaps related work, and Section~\ref{sec:conclusion} concludes this work.

\section{Background and Motivation}
\label{sec:background_motivation}
\minisection{Transformer Architecture.} 
The Transformer architecture~\cite{attention} has significantly advanced natural language processing (NLP) and has been foundational in large language models (LLMs).
A standard transformer model comprises an embedding layer to project sequences of tokens to hidden dimensions and stacks of transformer layers to capture long-term dependencies between input tokens using the self-attention mechanism.
The embedding layer is responsible for converting discrete words (tokens) into high-dimensional vectors that can be processed by neural networks, capturing both the semantic and positional information of individual tokens~\cite{word2vec}.
A transformer layer includes two main components: a multi-head attention (MHA) module and a position-wise feed-forward network (FFN), as shown in Figure~\ref{fig:background}(a).
The MHA module facilitates capturing contextual information by attending to different positions within the input sequence, while the FFN performs element-wise transformations to introduce non-linearity and improve representational capacity. 
 
At the core of transformer layers lies the attention module~\cite{attention}.
The relevant operations are given in Equation~\ref{eq:1} and~\ref{eq:2}.
The LayerNorm and residual connection are omitted for simplicity.
Three intermediate tensors are involved, namely, query $Q$, key $K$, and value $V$. 
The attention weights are calculated by first computing the dot product between $Q$ and $K$, then scaling the product by the square root of hidden dimension $d$ divided by the number of heads $h$, and finally going through a softmax operation ($\sigma(\cdot)$).
The attention scores $\text{Attn}(Q, K, V)$ are calculated by multiplying the attention weights by $V$. 
\begin{align}
    \text{Attn}(Q, K, V) = \sigma(\frac{QK^{T}}{\sqrt{d/h}}) \cdot V \label{eq:1}
\end{align}
The MHA output is obtained by simply concatenating the outputs of all attention heads along the head dimension, with each head being an attention module.
\begin{align}
    \text{MHA}(Q, K, V) = \text{Concat}(\text{Attn}_{1},...,\text{Attn}_{h})
    \label{eq:2}
\end{align}

\minisection{LLM Inference.} The generative inference of LLMs adopts an incremental decoding approach where the system computes the activations for all input prompt tokens in a single step and then iteratively decodes one new token using the input prompt and all previously generated tokens. 
Figure~\ref{fig:background}(b) gives an example of such autoregressive behavior in the LLM inference.
The inference procedure of the LLM inference can be divided into two parts, including the prefilling and decoding stages. 
As shown in the example, during the prefilling stage, LLMs first process all the input tokens in a single pass;
then, during the decoding stage, a previously generated output token is fed back into the model as an input and generates the next output token. 
Therefore, the decoding stage unfolds iteratively, processing one token at a time, until the sequence length reaches a maximum threshold or an ``$\langle$EOS$\rangle$'' token is emitted.

The distinctive autoregressive characteristics open up the opportunity of reusing intermediate states, specifically, the key ($K$) and value ($V$) tensors in the transformer layers, often referred to \textit{KV cache}~\cite{fairseq,efficientscaling}.
Note that the KV cache of one token depends on all previous tokens, and the size of the KV cache increases linearly as sequence length grows, as shown in Figure~\ref{fig:background}(c).
With the KV cache, the original quadratic-complexity calculation is replaced with linear-complexity memory accesses, thus significantly speeding up the LLM inference at the expense of additional GPU memory overhead.

\begin{figure}[!t]
\begin{center}
  \includegraphics[width=\linewidth]{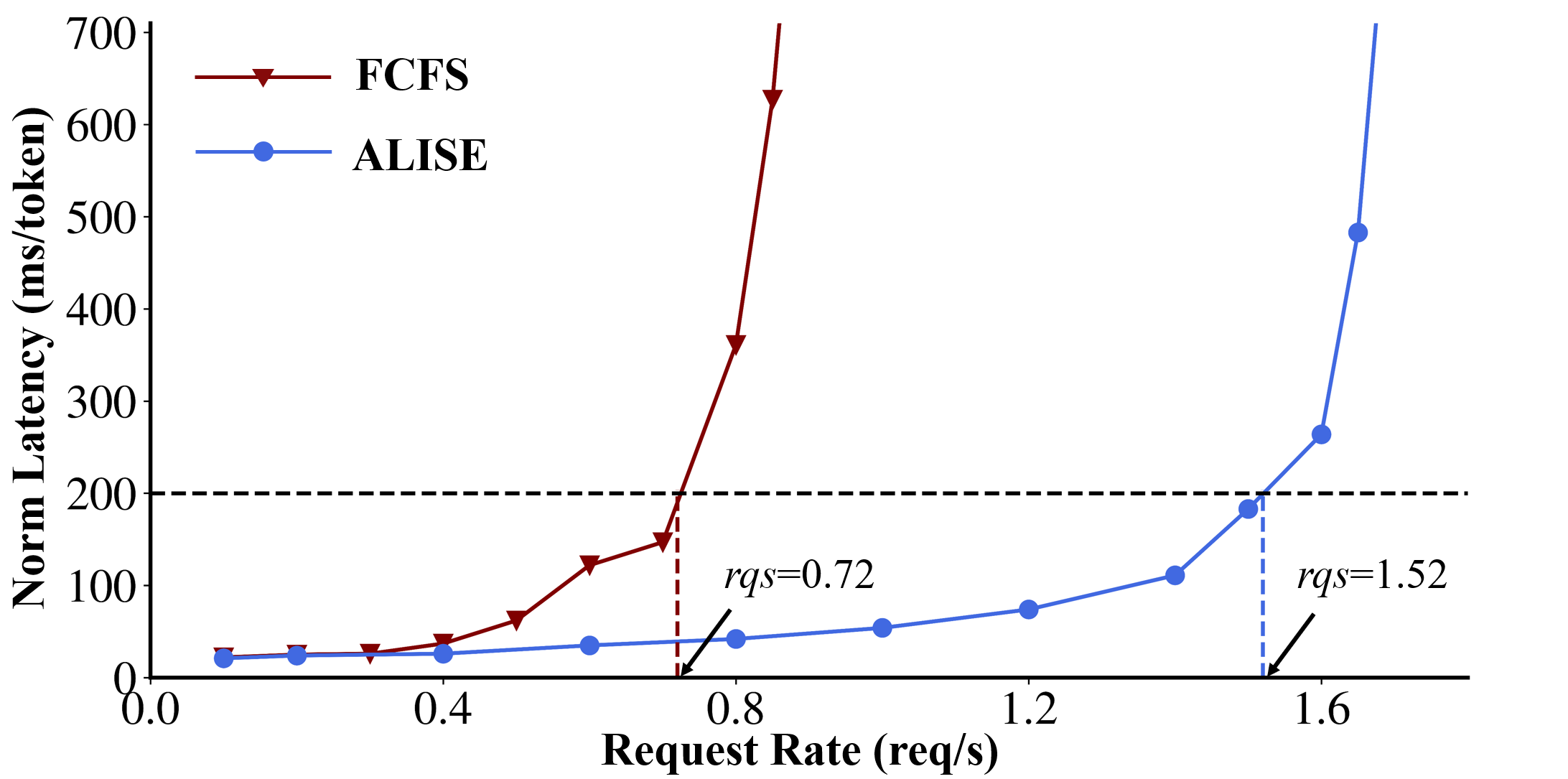}
\end{center}
\vspace{-4mm}
  \caption{
  End-to-end performance comparison of existing FCFS scheduling and speculative scheduling in ALISE on the ShareGPT dataset.
  }
\label{fig:motivation}
\end{figure}

\minisection{HoL Blocking in today's LLM Serving Systems.} 
In LLM inference, each token is generated based on the preceding ones, and this sequential dependency can result in unpredictable output lengths for generated sequences. 
The execution time for each request exhibits significant variability among input prompts, depending on both the input prompt and output length (number of iterations).
In FCFS, once a workload is scheduled, it must run until it finishes.
Run-to-completion scheduling is recognized for its head-of-line blocking issue, which is especially troublesome for generative inference tasks. 
This occurs when prior scheduled tasks, potentially lengthy in execution, hinder the timely processing of subsequent shorter tasks. 

To showcase such HoL blocking issues, we run the OPT-13B~\cite{opt} model on a single NVIDIA V100 GPU on the ShareGPT dataset~\cite{shareGPT} with varying sequence lengths to benchmark the baseline FCFS and \ourmethod.
Figure~\ref{fig:motivation} shows the end-to-end latency comparison of both methods.
As the request rate continually increases, FCFS induces much higher response latency due to inflexible non-preemptive scheduling.
In contrast, \ourmethod, with accurate and timely information about sequence length and execution time, can schedule the workloads appropriately, maximizing device usage, and providing up to $2\times$ better throughput under the same latency constraints.

\section{Methodology}
\label{sec:design}
\begin{figure}[!t]
\begin{center}
  \includegraphics[width=\linewidth]{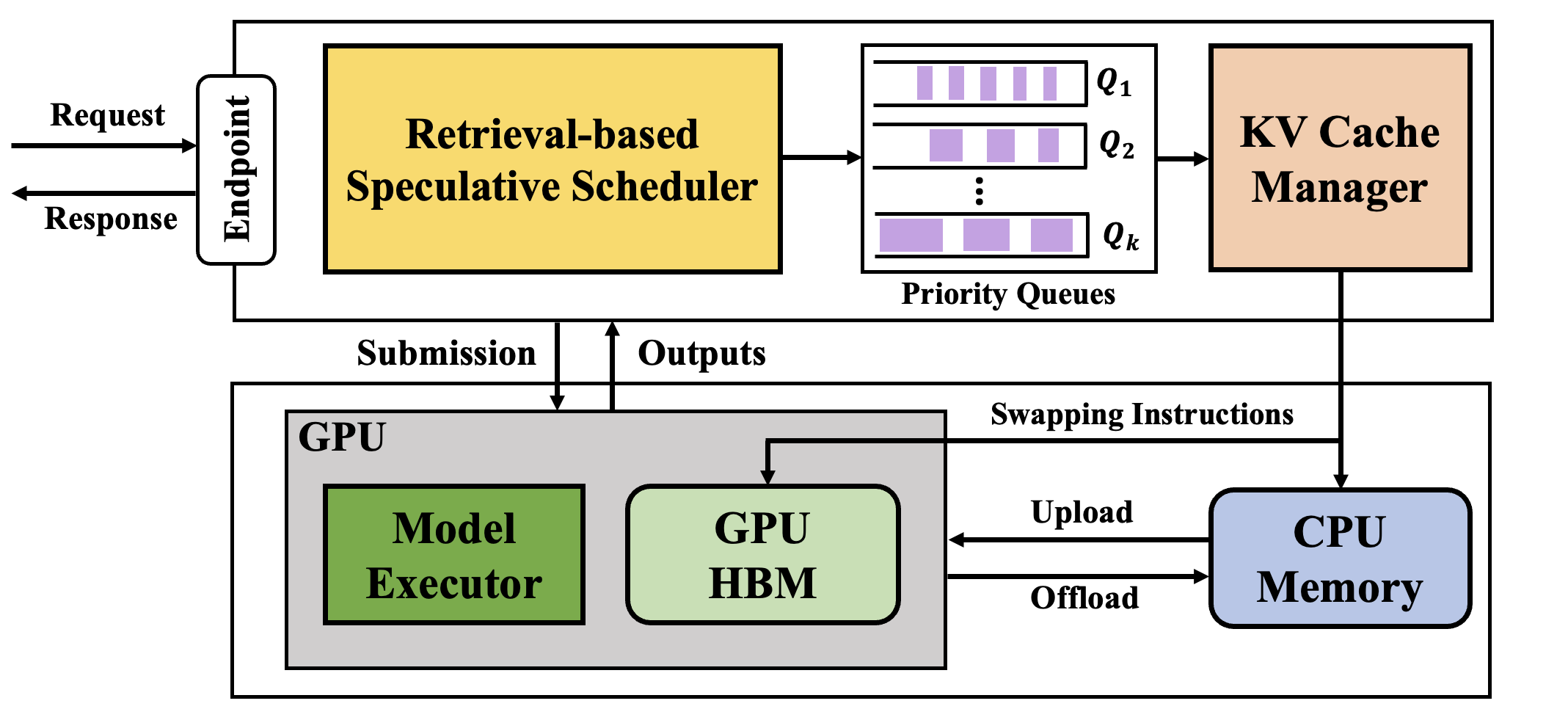}
\end{center}
\vspace{-4mm}
  \caption{System overview of \ourmethod. 
  }
\label{fig:arch}
\end{figure}

In this work, we develop a new speculative scheduling mechanism for LLM inference and build \ourmethod to tackle the challenges outlined in Section~\ref{sec:background_motivation}.
Figure~\ref{fig:arch} illustrates the architecture of \ourmethod. 
When users submit their requests to the inference service, the retrieval-based speculative scheduler first predicts the executive runtime and sends the job profile to priority queues to assign the appropriate priority level at the granularity of iteration, which favors the job with the shortest remaining time to address the HoL blocking issues. 
Once a job is selected, it will be automatically submitted to the GPU model executor, which retrieves the corresponding KV cache for computation.
\ourmethod also features an adaptive KV Cache manager that dynamically swaps the KV cache of preempted jobs between GPU HBM and CPU memory that overlaps with computation to avoid additional latency.

\subsection{Speculative Scheduler}
\begin{figure}[!t]
\begin{center}
  \includegraphics[width=\linewidth]{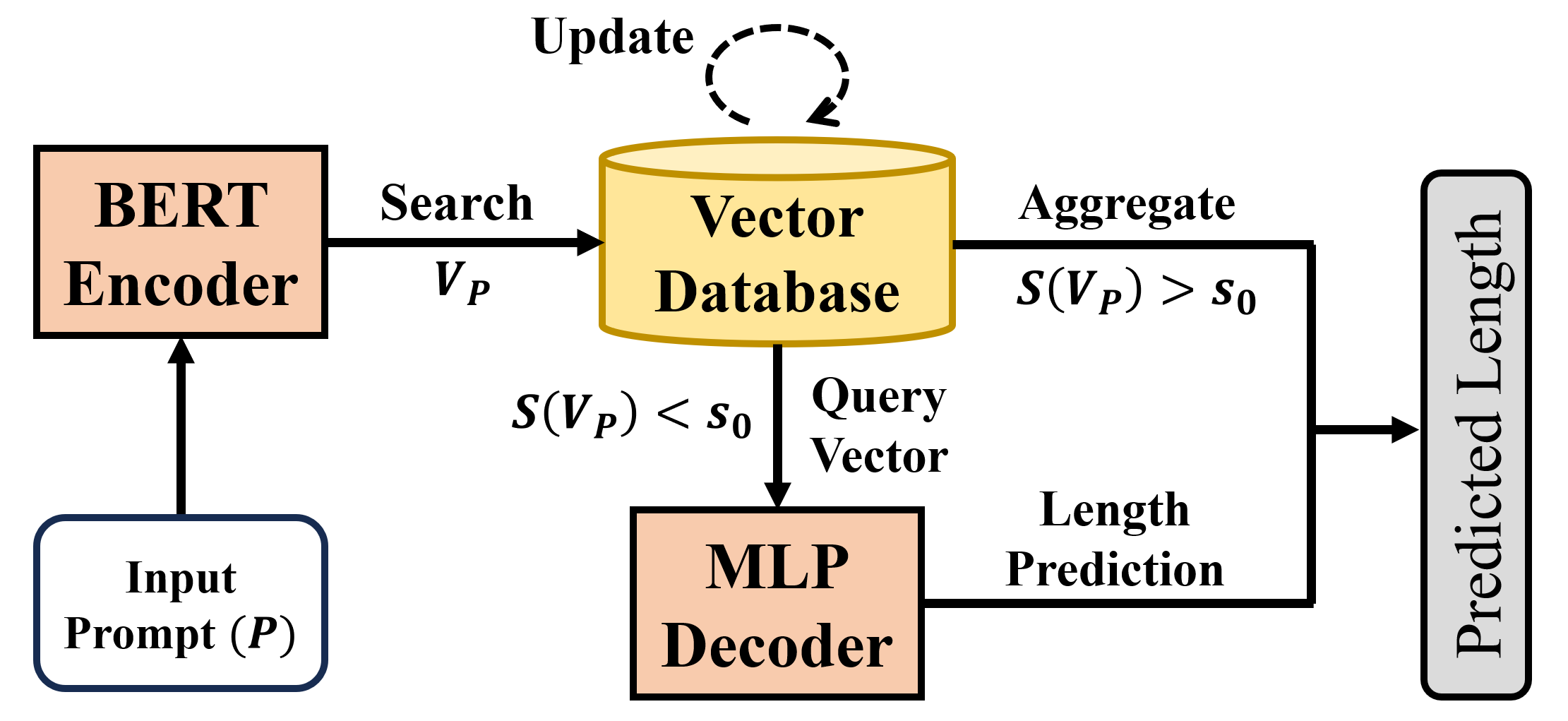}
\end{center}
\vspace{-4mm}
  \caption{Retrieval-based Length Predictor Architecture.}
\label{fig:re}
\end{figure}

\minisection{Output Length Prediction.} 
\begin{algorithm}[!b]
\begin{algorithmic}[1]
\Require Input prompts $\{P_{j}\}_{j=1,...,n}$, similarity threshold $s_0$, query database $DB$, search parameter $k$, BERT encoder $BERT\_{Enc}$ and all-MLP decoder $MLP\_{Dec}$.
\Ensure Predicted length $\{\hat{L}_{j}\}_{j=1,...,n}$
\While{true}
\For{all $j < n$} 
\State $V_{j} = BERT\_{Enc}(P_{j})$ \Comment{Embed input user prompt}
\State $I_{j} = \text{argmax}_{k} S(DB, V_{j})$ \Comment{Database vector search }
\If{$S(I_{j}) < s_0$} \Comment{Case I: No similar vectors found}
% \State \textcolor{gray}{\# Preemptive upload}
\State $\hat{L}_{j} = MLP\_{Dec}(I_{j})$ 
\Else \Comment{Case II: Similar vector(s) found}
\State $\hat{L}_{j} = \sum S(I_{j}) \cdot DB(I_{j})$ \Comment{Aggregate results}
\EndIf

\Return $\hat{L}_{j}$
\State $DB.update(V_{j},L_{j})$ \Comment{Update database}
\EndFor
\EndWhile
\end{algorithmic}
\caption{Retrieval-based Length Prediction}
\label{alg:re}
\end{algorithm}
A key aspect of enabling preemptive scheduling for LLM workloads is to predict the output sequence length.
Previous works have attempted to perform sequence prediction using smaller LLMs~\cite{it1,it2} or proxy models~\cite{s3,proxy}, however, they generally impose extensive engineering efforts and incur non-negligible overhead.
In this work, we propose to adopt a retrieval-based approach that leverages a vector database (DB) for estimating output length with high accuracy and low memory and latency overhead. Figure~\ref{fig:re} illustrates the details of our method.

We first transform the user prompt into high-dimension vectors using a pre-trained BERT encoder network~\cite{bert}. 
Next, we conduct a similarity search for input in the vector DB and retrieve top-$k$ historical queries with similarity scores above a pre-defined threshold.
The final output length is predicted using the weighted average of retrieved queries.
If no similar queries are found in the vector DB, the query context is fed to an all-MLP decoder fine-tuned for regression tasks to predict the sequence length.
After each request is finished, we update the vector DB with new queries and their actual response length to keep the dataset current and relevant. 
The details of this procedure are formulated as in Algorithm~\ref{alg:re}.

To evaluate the effectiveness of our retrieval-based method, we construct our database using the OpenChat~\cite{openchat} dataset and fine-tune the decoder on the Alpaca~\cite{alpaca} and ShareGPT~\cite{shareGPT} dataset.
Our approach achieves an average prediction error of 3.4\% and 9.2\% between predicted lengths and actual lengths on two datasets respectively, which is $1.4\times$ and $1.6\times$ lower than proxy-based methods.  
Moreover, thanks to the adoption of a query database, our method can reduce the prediction latency by up to 70\% by avoiding unnecessary computation.
This demonstrates that our speculative model can accurately predict output sequence length with low overhead.

\minisection{Predicting Execution Runtime.} Given the output length, we can thus formulate the execution of an inference job using a mathematical model. We first dissect the inference process into two parts, e.g. prefilling and decoding, and use the following to denote total execution time:
\begin{align}
    T_{gen}(s,n) = T_{pre}(s) + T_{dec}(s,n) \label{eq:3}
\end{align}
where $s$ denotes the input length, and $n$ denotes the output length. For the same hardware and model, the prefilling latency depends on the input sequence length, and the decoding latency is determined by both input and output length. 

\begin{figure}[!t]
\begin{center}
  \includegraphics[width=\linewidth]{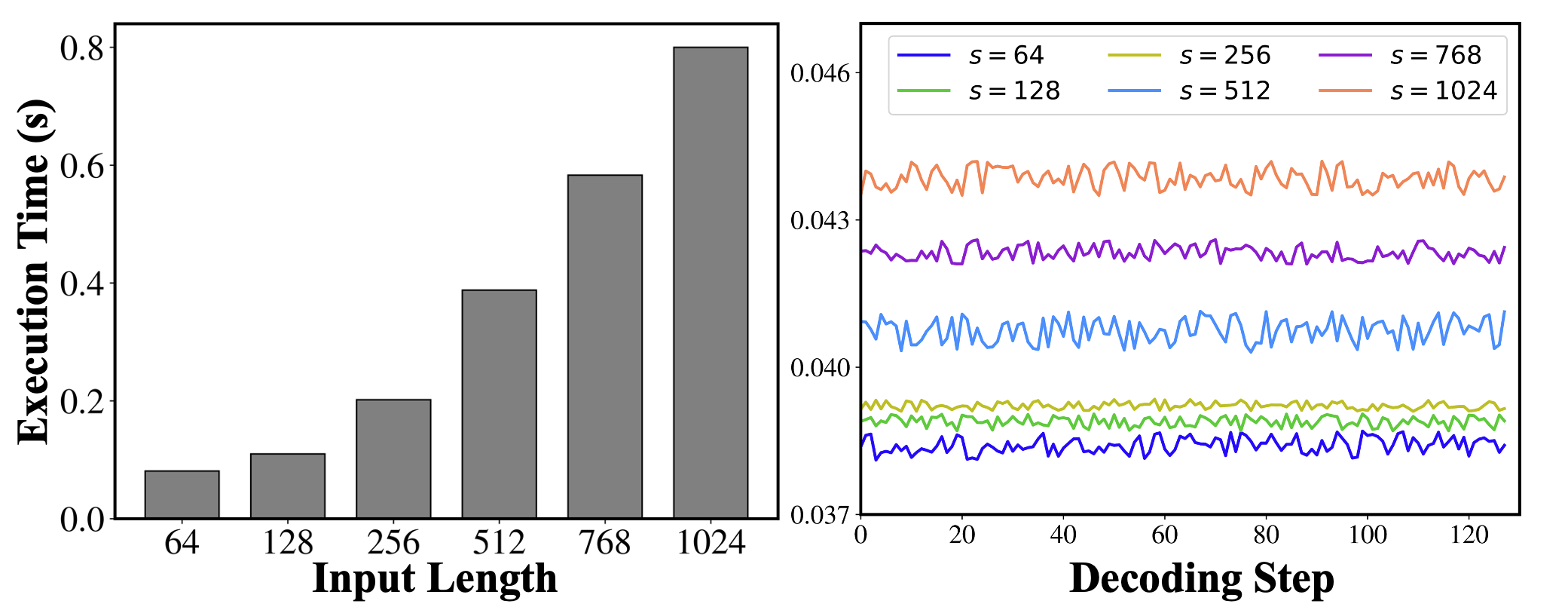}
\end{center}
\vspace{-4mm}
  \caption{
  Execution breakdown of the OPT-13B. 
  The left figure shows the prefill execution time with different input lengths ($s$), and the right figure shows the execution time for different decoding steps.}
\label{fig:breakdown}
\end{figure}

To further study the relationship between the sequence length and the execution latency of each stage, we perform an inference benchmark for OPT-13B under different input sequence settings.
As shown in Figure~\ref{fig:breakdown}, we can see that 1) the prefilling execution time $T_{pre}(s)$ increases almost linearly with input length $s$.
Coincidentally, the decoding latency exhibits a similar linear relationship concerning input length $s$, and the decoding latency $T_{dec}(s,n)$ for different iterations does not exhibit any noticeable changes, largely due to the KV cache. 
Based on these observations, we propose to estimate the prefilling and decoding latency as the following:
\begin{align}
     T_{pre}(s) &\approx s \cdot T_{0} \label{eq:4} \\
    T_{dec}(s,n)  &\approx n \cdot ( \alpha s + \beta) \label{eq:5}
\end{align}
where $\{T_{0}, \alpha, \beta\}$ are the linear regression coefficients that can be determined by profiling numerous samples of prefilling latency of different input and output sequence lengths.

\minisection{Preemptive Scheduling.} Leveraging the above analytical model that can accurately estimate the execution latency of LLM inference, we design a simple scheduler using priority queues~\cite{pq}.
The scheduler manages several job queues and assigns the priority order based on the estimated remaining execution time at the iteration level. 
We employ a virtual aging~\cite{va} mechanism for jobs in the lower priority queues to prevent starvation.
To handle short mispredictions, we demote the job to the lower priority queue if it exceeds its predicted length and doubles the predicted length.
With our proposed priority queues, our speculative scheduler can dynamically adjust the priority order for each job, and manage heterogeneous workloads more efficiently.
\subsection{Adaptive Memory Management}

Our proposed speculative scheduler provides iteration-level preemption to manage job scheduling orders, however, directly applying our methodology to LLM inference serving encounters the obstacle of additional memory overhead. 
In FCFS, we only need to store the intermediate KV cache for the executed jobs in the batch. In contrast, with preemptive scheduling, more jobs are in pending states in the GPU, and additional memory space must be allocated to store their KV cache for future use.
Determined by the model size, batch size, and sequence length, the KV cache could consume significant memory space, potentially causing out-of-memory (OOM) errors, and halting the execution~\cite{flexgen,alisa}. 

A straightforward solution to mitigate the memory overhead issues is to defer all the newly arrived jobs in the priority queue when the GPU memory reaches the maximum capacity. 
However, this solution loses its validity when incoming new jobs have higher priority (shorter execution time), but are blocked due to memory space limitation. 
In resource-constrained settings, such as single GPU-CPU systems, our speculative scheduling would be downgraded to FCFS, and the HoL blocking issues would resurface. 
Another practical solution is to perform deletion for the unused KV cache for jobs in the queues to make room for the new jobs. 
But this solution suffers from two major downsides: 1) the deleted KV cache has to be recomputed when low-priority jobs are scheduled for execution, which induces additional compute latency; 
2) with the aging mechanism, the killed low-priority jobs would be promoted to the high-priority queue after passing the age threshold, which then kills the currently executed jobs, leading to potential deadlocks.

\minisection{Dynamic Swapping.} 
Previous works have explored leveraging the memory of different hierarchies to accelerate LLM inference in resource-constrained environments, however, they are mostly for offline batched inference and schedule the KV cache swapping at either head or token level~\cite{flexgen,alisa}.
In this work, we design a priority-based KV cache manager that dynamically swaps the KV cache for preempted jobs between GPU HBM and CPU memory at the request level. 
Specifically, \ourmethod performs dynamic swapping operations of the KV cache based on the metric of estimated wait time (EWT).
Generally, jobs with higher EWT tend to be offloaded to the CPU DRAM to make space for high-priority jobs, while jobs with lower EWT tend to be uploaded for later execution. 
Additionally, the offloading and uploading operations are overlapped with computation to avoid additional latency.

For a job with priority $p$, we can calculate the EWT by summing all the execution time of jobs with higher priority as follows:
\begin{align}
    EWT(J) = \sum_{m < p}^{m} Predictor(J) \label{eq:ewt}
\end{align}
Due to the existence of the aging mechanism (age threshold $K$), we also need to compare the time needed to promote $J_{i}$ to the high-priority queue for execution. Therefore, we reformulate Equation~\ref{eq:ewt} as follows:
\begin{align}
    EWT(J) = \min (\sum_{m < p}^{m} Predictor(J), T_{promote}(J, K))
\end{align}

\begin{algorithm}[!t]
\begin{algorithmic}[1]
\Require Priority job queue $\{Q_{i}\}_{i=1,...,h}$, GPU memory $GM$, CPU memory $CM$, GPU job limit $M$.
\While{true}
\For{$q \in \{Q_1, ..., Q_h\}$}
\State $\text{EWT}(q).sort()$
\For{$i=1$ \textbf{to} $len(q)$}
\If{$J_{i}$ not in GPU \textbf{and} $i < M$}
\State $CM$.upload($J_{i}$) \Comment{Preemptive upload}
\Else
\State $GM$.offload($J_{i}$) \Comment{Preemptive offload}
\EndIf
\If{$len(q) < M$} \Comment{Update GPU job limit}
\State $M = M - len(q)$
\EndIf
\EndFor
\EndFor
\EndWhile
\end{algorithmic}
\caption{Dynamic Swapping}
\label{alg:da}
\end{algorithm}

Leveraging the above EWT definitions, we can decide the order of offloading and uploading of each job by setting the maximum number of jobs allowed in the GPU, which is constrained by the total available GPU memory space.
The details of our dynamic swapping is shown in Algorithm~\ref{alg:da}. 
For each priority queue, we first calculate and sort the EWT of all the jobs in the queue (lines 2-3).
We keep the preempted jobs in the GPU if not exceeding the GPU job limit $M$, and perform the preemptive upload of the KV cache if not in the GPU (lines 5-6).
For jobs exceeding the job limit, we perform preemptive offload operations (lines 7-8).
The GPU limit $M$ is also updated by each queue accounting for the size of previous queues (lines 10-11).

\minisection{KV Compression.}
Previous works have utilized quantization to accelerate attention computation by compressing model weights~\cite{GPTQ,AWQ}.
In this work, we leverage quantization for a different purpose, i.e., compressing the unused KV cache to reduce memory overhead. 
We adopt a fine-grained channel-wise quantization for the KV cache to ensure robust performance and minimal information loss. 
More specifically, we use the following formula to quantize KV cache to $b$-bit integers in memory and de-quantize them to their original format (FP16 in this work) for computation:
\begin{align}
    x_{\text{q}} = \text{round}(\frac{1}{\lambda}x + z), \quad x= \lambda(x_{\text{q}} - z) 
\end{align}
where zero point $z= \nint{(\frac{-2^b}{max - min})}$, and the scaling factor $\lambda=\frac{max - min}{2^b - 1}$.
In this work, we choose the 8-bit integer for quantizing the KV cache to ensure minimal degradation on text generation quality.

\subsection{Implementation}
\ourmethod is implemented on top of PyTorch~\cite{pytorch} and HuggingFace Transformer libraries~\cite{huggingface} with 4.5K lines of Python.
Our execution engine is based on vLLM~\cite{vLLM}, which we modify to support iteration-level preemptive scheduling and interact with the KV cache manager for dynamic swapping. 
\ourmethod also features customized fused kernels for LayNorm, Attention, and ReLU operations, just like other systems for transformer-based LLMs~\cite{vLLM,orca,fastertransformer}. 
We implement these kernels based on python-based Triton compiler~\cite{openaitriton}.
To increase the compute utilization of GPUs and improve inference efficiency, we adopt the iteration-level continuous batching technique~\cite{orca}, where we apply batching only to a handful of operations, such as Linear, LayerNorm, and ReLU, which are compatible with irregularly shaped tensors.

\begin{figure*}[!t]
\begin{center}
  \includegraphics[width=\linewidth]{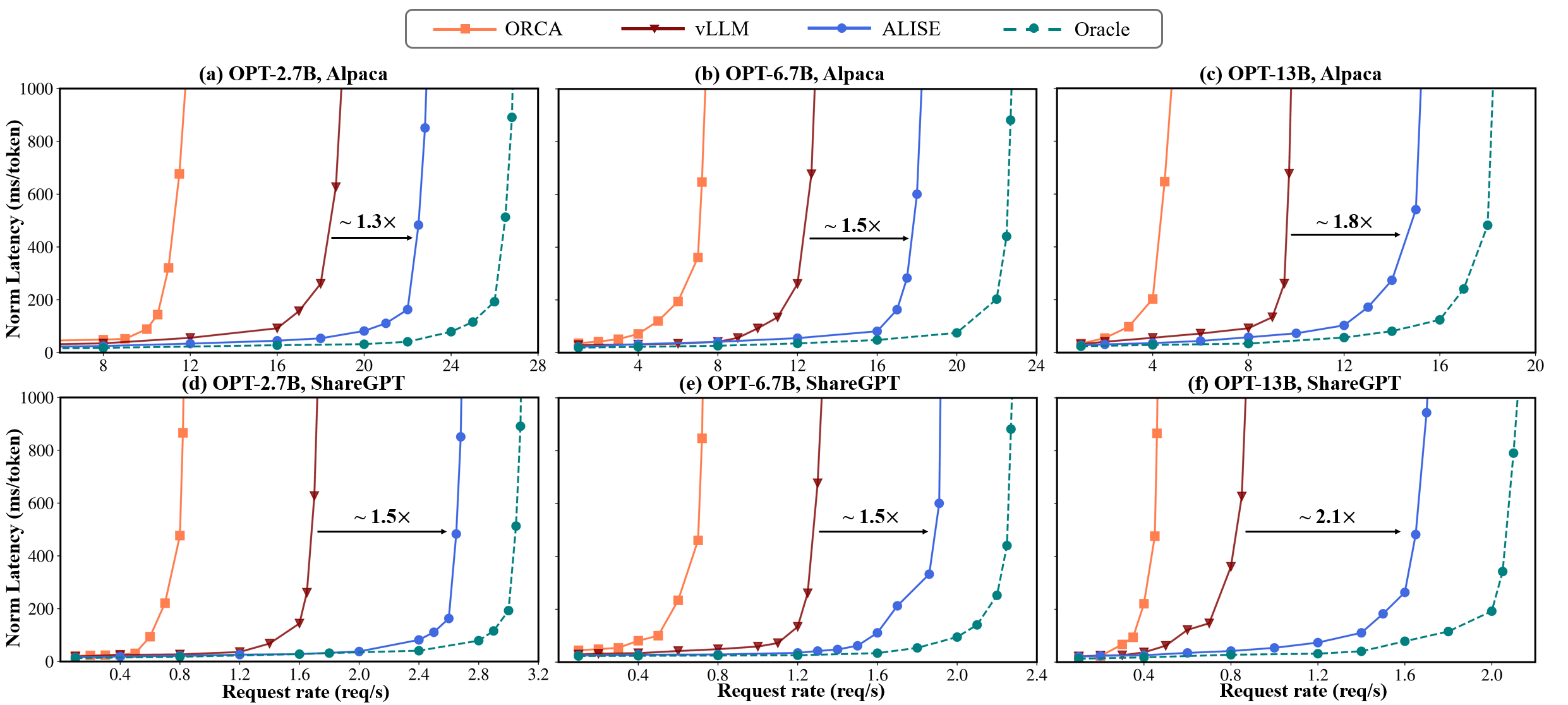}
\end{center}
\vspace{-2mm}
  \caption{
  End-to-end performance comparison of \ourmethod and baselines, including ORCA~\cite{orca}, vLLM~\cite{vLLM}, and the Oracle on the Alpaca~\cite{alpaca} and ShareGPT dataset~\cite{shareGPT}.
  }
\label{fig:main}
\end{figure*}

\section{Evaluation}
\label{sec:evaluation}
\subsection{Experimental Setup}
\minisection{Hardware.} We conduct our experiments on a single customized high-performance GPU instance, configured with two NVIDIA Tesla V100 GPUs with 32\,GB VRAM, and a 2.60\,GHz Intel Xeon CPU connected over PCIe 4.0$\times$16 and 1024\,GB DRAM.

\begin{figure}[!t]
\begin{center}
  \includegraphics[width=\linewidth]{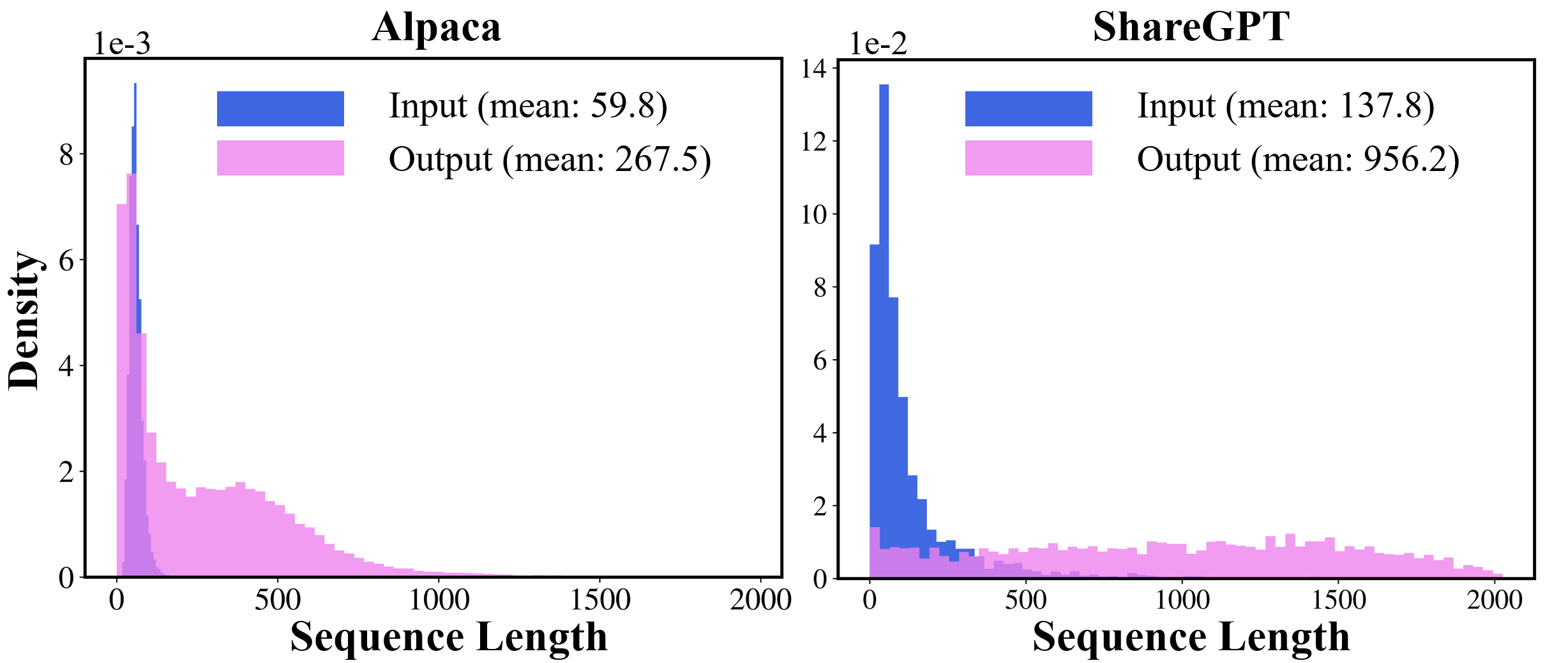}
\end{center}
\vspace{-4mm}
  \caption{Input and output length distribution of the Alpaca and ShareGPT datasets.}
\label{fig:dist}
\end{figure}

\minisection{Workloads.} 
To emulate the LLM workloads, we synthesize traces of user requests based on the Alpaca~\cite{alpaca} and ShareGPT~\cite{shareGPT} dataset since there is no publicly available request trace for LLMs. 
The Alpaca dataset is a GPT-3.5 generated instruction dataset and is widely used for LLM inference evaluation~\cite{it1,vLLM}. 
The ShareGPT dataset is a public dataset that collects the chatbot conversation from ChatGPT users, which exhibit greater variance in contents and sequence length, as shown in Figure~\ref{fig:dist}.
We tokenize each dataset and use their input and output lengths to simulate real-world user requests.
In terms of request timestamps, we generate the arrival time based on the Poisson distribution.

\minisection{Models.} Following prior works on LLM serving, we use the popular OPT~\cite{opt}, one of the most representative open-sourced LLM families widely used in both industry and academia, for our main evaluation.
We choose three model size configurations, namely 2.7B, 6.7B, and 13B.
Table~\ref{tab:1} lists the detailed model and corresponding GPU configurations. 
For all experiments, we maintain the models in the GPU HBM during serving and use FP16 precision for model weights, activations, and INT8 for the KV cache. 

\minisection{Baselines.}
We compare \ourmethod with three baselines.
\begin{itemize}
    \item \textbf{ORCA}~\cite{orca}: ORCA is one of the first LLM systems to adopt the iteration-level first-come-first-serve (FCFS) scheduling at the iteration level. 
    \item \textbf{vLLM}~\cite{vLLM}: vLLM is the state-of-the-art online LLM inference serving system, which manages the KV cache at the block level but still uses FCFS protocol. 
    \item \textbf{Oracle}: We also implement an ideal system baseline with a perfect predictor that knows the output lengths, which we term Oracle, to show the upper bound of our system performance. 
\end{itemize}

\minisection{Metrics.} To measure the inference serving throughput, we use the normalized latency of the systems, defined as the mean of the request's end-to-end latency divided by the number of generated tokens.
A high-throughput serving system should retain low normalized latency against high request rates.
We set the batch size as large as possible according to the GPU memory capacity.
All system performances are evaluated with 30-minute traces.

\begin{table}[!t]
\begin{center}
\caption{Model configurations.}
\vspace{-4mm}
\resizebox{\columnwidth}{!}{
    \begin{tabular}{lcccc}
    \toprule
    \textbf{Model} & \# Heads & \# Layers &  Hidden Size & Parameter Size \\ \midrule
    OPT-2.7B & 32 & 32 & 2560 & 5\,GB \\ 
    OPT-6.7B & 40 & 40 & 5120 & 13\,GB\\ 
    OPT-13B & 40 & 40 & 5120 & 24\,GB\\ 
    \bottomrule
    \end{tabular}
\label{tab:1}
}
\end{center}
\end{table}

\begin{figure*}[!t]
\begin{center}
  \includegraphics[width=\linewidth]{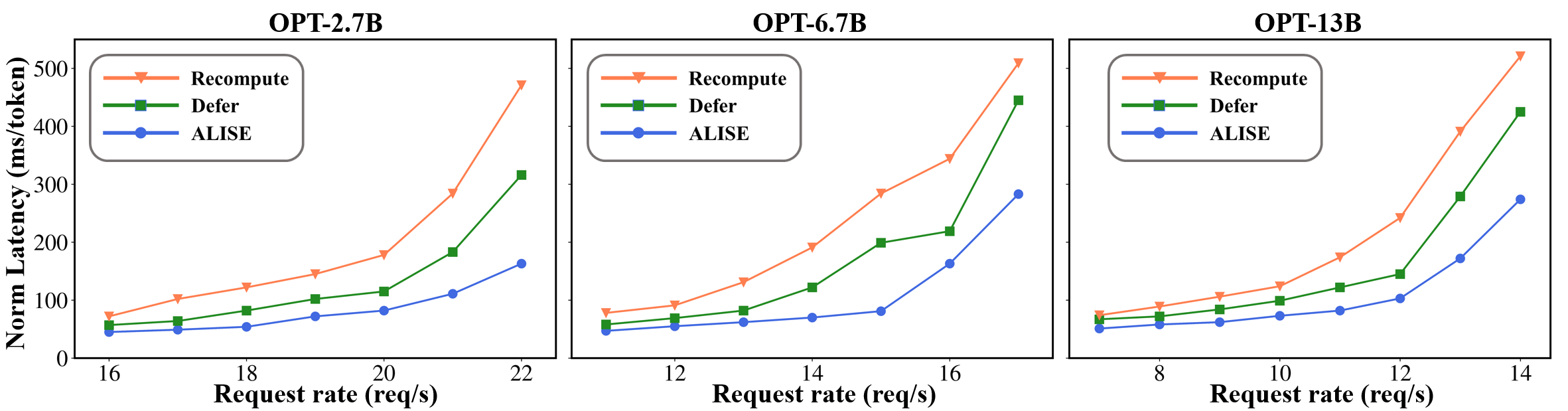}
\end{center}
\vspace{-4mm}
  \caption{
  Latency comparison of different memory management protocols in \ourmethod on the Alpaca dataset.  
  }
\label{fig:kv_ab}
\end{figure*}

\begin{table*}[!t]
\begin{center}
\caption{Accuracy and throughput evaluation of our proposed retrieval-based length predictor on the ShareGPT dataset.}
\vspace{-3mm}
\resizebox{2\columnwidth}{!}{
\begin{tabular}{r|cc|cc|cc}
\toprule
\multirow{2}{*}{Metrics} & \multicolumn{2}{c|}{OPT-2.7B} & \multicolumn{2}{c|}{OPT-6.7B} & \multicolumn{2}{c}{OPT-13B} \\ \cline{2-7} 
 & Proxy-based & Retrieval-based & Proxy-based & Retrieval-based & Proxy-based & Retrieval-based \\ \hline
Accuracy ($\uparrow$) & 0.781 & \textbf{0.821} & 0.712 & \textbf{0.856} & 0.634 & \textbf{0.744} \\
Pred. Error ($\downarrow$) & 0.122 & \textbf{0.057} & 0.145 & \textbf{0.096} & 0.178 & \textbf{0.123} \\
Avg. Pred. Latency ($\downarrow$) & 12.2 ms & \textbf{3.92 ms} & 11.7 ms & \textbf{4.74 ms} & 14.8 ms & \textbf{4.49 ms} \\
Throughput ($\uparrow$) & $1\times$ & \textbf{$1.47\times$} & $1\times$ & \textbf{$1.63\times$} & $1\times$ & \textbf{$1.82\times$} \\ 
\bottomrule
\end{tabular}
\label{tab:3}
}
\end{center}
\end{table*}

\subsection{End-to-End Performance}
The first row of Figure~\ref{fig:main} shows the results on the Alpaca dataset. We can see that as the request rate increases, the normalized latency gradually increases but then explodes after a certain threshold. 
Such observation aligns with previous works~\cite{orca,vLLM,flexgen, alisa}, where LLM inference is often \textit{memory-bound}, and once the request rate reaches the memory capacity of the serving system, the response latency grows infinitely as the request queue pool continues to expand.
On the Alpaca dataset, \ourmethod obtains $1.3\sim1.8\times$ higher throughput compared to vLLM while maintaining similar latency results.
This is due to \ourmethod's speculative scheduling strategy that dynamically adjusts the priority order of each request thus reducing the response delay.
For instance, as shown in Figure~\ref{fig:main}(b), under the latency constraints of 200 ms, \ourmethod processes $1.5\times$ more requests than vLLM.
When compared against ORCA, \ourmethod sustains up to $3.1\times$ higher request rates.
The advantage of \ourmethod is more pronounced on the ShareGPT dataset, as shown in the second row of Figure~\ref{fig:main}. 
This is because the ShareGPT dataset contains longer input prompts and outputs on average than the Alpaca dataset, with much higher variance.
\ourmethod achieves up to $2.1\times$ throughput improvement against the state-of-the-art vLLM.
Compared to ORCA, \ourmethod also obtains up to $4.3\times$ high throughput under similar latency constraints.

\begin{figure}[!t]
\begin{center}
  \includegraphics[width=\linewidth]{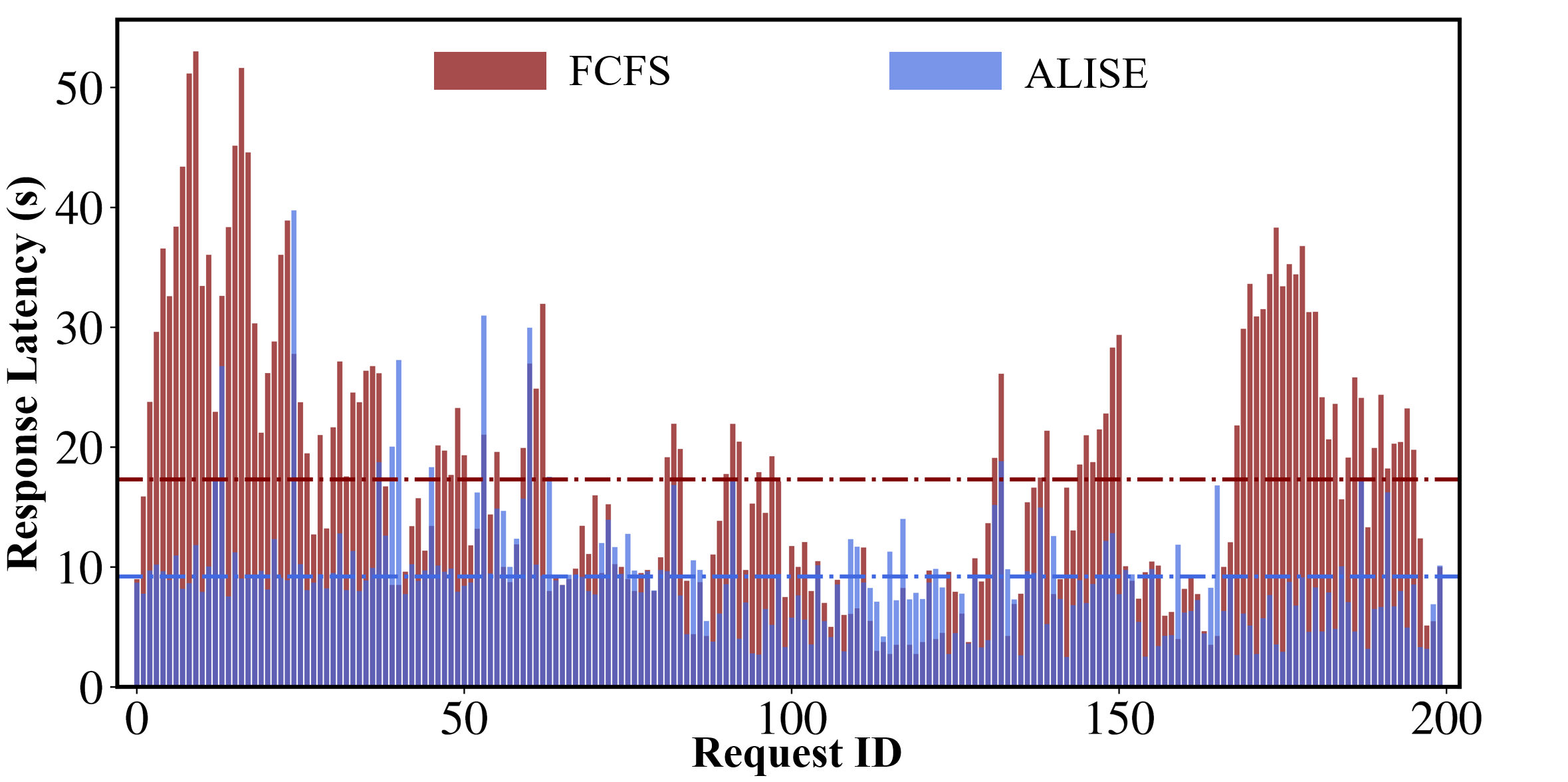}
\end{center}
\vspace{-4mm}
  \caption{
  Response latency for 200 sampled requests when serving OPT-13B on the ShareGPT dataset at 2 req/s. 
  The dotted line indicates the mean response latency.}
\label{fig:mb}
\end{figure}

\subsection{Performance Analysis}
\minisection{Impact of KV Cache Management.} We further demonstrate the benefits of our proposed adaptive KV cache management by comparing our design with two strawman solutions, namely \textit{Recompute} and \textit{Defer} strategy. 
The \textit{Recompute} strategy deletes the KV cache of preempted low-priority jobs and recomputes the KV cache when jobs are rescheduled for execution, while the \textit{Defer} strategy simply defers newly arrived jobs when the GPU does not have enough memory space for executing the job.

We perform the experiments on the OPT models with varying request rates on the Alpaca dataset. 
As shown in Figure~\ref{fig:kv_ab}, when the request rate is relatively low, the gap between \ourmethod and the other two methods is not significant, as the KV cache for most jobs is in the GPU.
However, as the request rate increases, the gap between these solutions is widened:
\ourmethod outperforms \textit{Defer} and \textit{Recompute} by up to $1.8\times$ and $2.8\times$ in terms of normalized latency under the same request rate.
For the \textit{Defer} strategy, with a higher request rate, the HoL blocking issues resurface, as more newly arrived jobs with shorter execution times are delayed, thus leading to increased response latency;
for the \textit{Recompute} strategy, the increased latency is largely attributed to the increased execution time for token generation, since for scheduled preempted jobs, the system has to recompute the KV cache of previous iterations to generate the subsequent tokens.

\minisection{Impact of Length Predictor.}
To further show the effectiveness of our proposed retrieval-based length predictor, we conduct experiments to compare the accuracy and throughput performance with previous proxy-based predictors~\cite{s3,proxy} on the ShareGPT dataset.
To measure the prediction accuracy, we categorize the lengths into evenly divided bins, and we regard the prediction as accurate when both lengths fall into the same bucket.
In Table~\ref{tab:3}, we can see that our method improves previous methods in both aspects. 
Specifically, thanks to the introduction of a query database storing history user information, a retrieval-based approach can predict new queries with high accuracy and low prediction error;
as some queries do not need to go through the predictor model, our method also achieves significant speedup against proxy-based methods regarding prediction latency.
Moreover, with more accurate length information, the scheduler can assign more appropriate priority orders accordingly, improving the overall inference throughput performance.

\minisection{Response Latency Analysis.}
To better understand our speculative scheduler on the system throughput, we randomly sample 200 consecutive generated requests during serving OPT-13B on the ShareGPT dataset and compare the end-to-end response latency of each request when served with non-preemptive FCFS scheduling and \ourmethod.
As shown in Figure~\ref{fig:mb}, our system significantly reduces the response latency for most requests, thanks to the preemption of speculative scheduling, which alleviates the head-of-line blocking issues.
\ourmethod also achieves close to 46\% reduction in mean response latency against FCFS baselines.

\minisection{Results on More Models.} To verify the effectiveness of our method on different models, we choose several other popular open-sourced LLMs, such as LLaMA-7B/13B~\cite{llama2} and Pythia-12B~\cite{pythia} for further evaluation.
As shown in Table~\ref{tab:2}, \ourmethod provides consistent improvement over previous baseline systems for these LLMs.
\begin{table}[!t]
\begin{center}
\caption{Throughput improvement for different models on the ShareGPT and Alpaca datasets, measured in processed requests per second.}
\vspace{-3mm}
% \resizebox{0.82\columnwidth}{!}{
\begin{tabular}{lccc}
\toprule
\multicolumn{1}{l|}{} & \multicolumn{1}{c|}{ORCA} & \multicolumn{1}{c|}{vLLM} & ALISE \\ 
\midrule
\multicolumn{4}{l}{Alpaca (30 req/s)} \\ 
\midrule
\multicolumn{1}{l|}{LLaMA-13B} & \multicolumn{1}{c|}{42.23} & \multicolumn{1}{c|}{71.87} & \multicolumn{1}{c}{\textbf{101.42} ($+41\%$)} \\
\multicolumn{1}{l|}{LLaMA-7B} & \multicolumn{1}{c|}{75.42} & \multicolumn{1}{c|}{122.87} & \multicolumn{1}{c}{\textbf{164.51} ($+34\%$)} \\ 
\multicolumn{1}{l|}{Pythia-12B} & \multicolumn{1}{c|}{34.85} & \multicolumn{1}{c|}{64.19} & \multicolumn{1}{c}{\textbf{91.12} ($+42\%$)} \\ 
\midrule
\multicolumn{4}{l}{ShareGPT (2 req/s)} \\ 
\midrule
\multicolumn{1}{l|}{LLaMA-13B} & \multicolumn{1}{c|}{14.42} & \multicolumn{1}{c|}{27.89} & \multicolumn{1}{c}{\textbf{41.22} ($+47\%$)} \\
\multicolumn{1}{l|}{LLaMA-7B} & \multicolumn{1}{c|}{30.28} & \multicolumn{1}{c|}{62.93} & \multicolumn{1}{c}{\textbf{87.32} ($+39\%$)} \\
\multicolumn{1}{l|}{Pythia-12B} & \multicolumn{1}{c|}{12.36} & \multicolumn{1}{c|}{24.95} & \multicolumn{1}{c}{\textbf{37.27} ($+49\%$)}\\ 
\bottomrule
\end{tabular}
\label{tab:2}
% }
\end{center}
\end{table}
\section{Related Work}
\label{sec:related_work}
\minisection{DNN Inference Serving.}
Recent years have witnessed the exponential of AI-driven applications in real-world practices, and numerous dedicated systems~\cite{clipper,clockwork,shepherd,infaas} have been developed to meet such growing demand.
They generally focus on batching~\cite{clipper}, scheduling~\cite{clockwork}, and caching optimizations~\cite{shepherd} for serving small-scale models such as ResNet.
More recently, INFaaS~\cite{infaas} automates the model selection and deployment process by navigating the best trade-off between model variants and performances.
However, these systems fail to take into account the multi-iteration characteristics and the unpredictability of LLM inference, resulting in missed opportunities for further optimizations.

\minisection{LLM Inference Serving.}
Several specialized inference serving systems have been developed for Transformer-based LLMs~\cite{fastertransformer,orca,vLLM}.
FasterTransformer is among the first to optimize training and inference serving for large transformer-based LLMs by utilizing GPU kernel optimizations and model parallelism~\cite{fastertransformer}. 
ORCA further develops iteration-level scheduling and selective batching by leveraging the multi-iteration nature of LLM inference~\cite{orca}.
vLLM proposes PagedAttention to store the KV cache in a non-contiguous paged memory to alleviate memory fragmentation~\cite{vLLM}.
While they have achieved promising results, existing LLM systems use the non-preemptive first-come-first-served (FCFS)~\cite{fcfs} to process LLM inference workloads. 
It is known that such run-to-completion scheduling exhibits head-of-line blocking since once a workload is scheduled, it must run until it finishes. An LLM inference job with a long output length would run for a long time to block following relatively short jobs, thus causing head-of-line blocking and increasing the response latency.
Given the heterogeneous nature of real-world requests, the inflexible FCFS scheduling has become the new bottleneck for low-latency and high-throughput LLM serving systems.

\minisection{LLM Output Length Prediction.}
There are several limited works on predicting the LLM output sequence length.
$S^3$ adopts a DistillBERT~\cite{distillbert} model to classify the output length range given LLM input queries, while SSJF~\cite{proxy} employs a similar proxy model to perform classification and regression for sequence length.
Instruction tuning has also been leveraged to prompt LLM itself to output sequence length, but it generally introduces additional bias and requires intensive engineering efforts~\cite{it1,it2}.
In our work, we develop a retrieval-based predictor building upon a vector database, which provides more accurate length prediction at low latency overhead.

\minisection{Memory Optimization for LLMs.} 
Numerous techniques have been proposed to reduce the memory overhead of LLM inference. 
SparseTransformer accelerates the inference process by generating sparsity patterns with a fixed stride on all sequence tokens~\cite{sparseformer}.
FlashAttention aims to reduce memory accesses between on-chip SRAM and off-chip HBM in GPUs with fine-grained tiling and partitioning at the kernel level~\cite{flashattention, flashattention2}.
FlexGen focuses on resource-constrained single GPU-CPU scenarios and formulates a static offloading strategy for KV cache, model weights, and intermediate results utilizing CPU DRAM and secondary storage~\cite{flexgen}. 
ALISA proposes an algorithm-system co-design approach that exploits the sparsity patterns in LLM inference to achieve the optimal trade-offs between caching and recomputation~\cite{alisa}. 
Orthogonal to existing memory optimization techniques, our work focuses on scheduling optimization to minimize response latency for serving online heterogeneous workloads.

\section{Conclusion}
\label{sec:conclusion}
In this work, we present a new efficient LLM inference serving solution, termed \ourmethod, to address the head-of-line (HoL) blocking issues in the existing FCFS scheduling strategy.
\ourmethod leverages a novel retrieval-based speculative to schedule incoming jobs by their remaining execution time preemptively, thus minimizing potential queuing delays for heterogeneous workloads.
To alleviate the memory overhead of preempted jobs, we design adaptive memory management that dynamically performs swapping operations for unused KV cache and quantization-based compression.
Experiments demonstrate that \ourmethod obtains up to $1.8\times$ and $2.1\times$ throughput improvement over the state-of-the-art systems, such as vLLM, on the Alpaca and ShareGPT datasets, respectively. 

\section*{Acknowledgement}
\label{sec:ack} 
This work was sponsored in part by the U.S. National Science Foundation (NSF) under Grants 1907765 and 2400014.
The authors would like to thank the anonymous ICCAD reviewers for their constructive feedback to improve this work.
\bibliographystyle{ACM-Reference-Format}
\bibliography{sample-base}

%%% -*-BibTeX-*-
%%% Do NOT edit. File created by BibTeX with style
%%% ACM-Reference-Format-Journals [18-Jan-2012].

\begin{thebibliography}{40}

%%% ====================================================================
%%% NOTE TO THE USER: you can override these defaults by providing
%%% customized versions of any of these macros before the \bibliography
%%% command.  Each of them MUST provide its own final punctuation,
%%% except for \shownote{}, \showDOI{}, and \showURL{}.  The latter two
%%% do not use final punctuation, in order to avoid confusing it with
%%% the Web address.
%%%
%%% To suppress output of a particular field, define its macro to expand
%%% to an empty string, or better, \unskip, like this:
%%%
%%% \newcommand{\showDOI}[1]{\unskip}   % LaTeX syntax
%%%
%%% \def \showDOI #1{\unskip}           % plain TeX syntax
%%%
%%% ====================================================================

\ifx \showCODEN    \undefined \def \showCODEN     #1{\unskip}     \fi
\ifx \showDOI      \undefined \def \showDOI       #1{#1}\fi
\ifx \showISBNx    \undefined \def \showISBNx     #1{\unskip}     \fi
\ifx \showISBNxiii \undefined \def \showISBNxiii  #1{\unskip}     \fi
\ifx \showISSN     \undefined \def \showISSN      #1{\unskip}     \fi
\ifx \showLCCN     \undefined \def \showLCCN      #1{\unskip}     \fi
\ifx \shownote     \undefined \def \shownote      #1{#1}          \fi
\ifx \showarticletitle \undefined \def \showarticletitle #1{#1}   \fi
\ifx \showURL      \undefined \def \showURL       {\relax}        \fi
% The following commands are used for tagged output and should be
% invisible to TeX
\providecommand\bibfield[2]{#2}
\providecommand\bibinfo[2]{#2}
\providecommand\natexlab[1]{#1}
\providecommand\showeprint[2][]{arXiv:#2}

\bibitem[AI(2024)]%
        {llama3}
\bibfield{author}{\bibinfo{person}{Meta AI}.} \bibinfo{year}{2024}\natexlab{}.
\newblock \bibinfo{title}{Introducing Llama 3.1: Our most capable models to date}.
\newblock
\newblock
\urldef\tempurl%
\url{https://ai.meta.com/blog/meta-llama-3-1/}
\showURL{%
\tempurl}


\bibitem[Biderman et~al\mbox{.}(2023)]%
        {pythia}
\bibfield{author}{\bibinfo{person}{Stella~Rose Biderman}, \bibinfo{person}{Hailey Schoelkopf}, \bibinfo{person}{Quentin~G. Anthony}, \bibinfo{person}{Herbie Bradley}, \bibinfo{person}{Kyle O'Brien}, \bibinfo{person}{Eric Hallahan}, \bibinfo{person}{Mohammad~Aflah Khan}, \bibinfo{person}{Shivanshu Purohit}, \bibinfo{person}{Usvsn~Sai Prashanth}, \bibinfo{person}{Edward Raff}, \bibinfo{person}{Aviya Skowron}, \bibinfo{person}{Lintang Sutawika}, {and} \bibinfo{person}{Oskar van~der Wal}.} \bibinfo{year}{2023}\natexlab{}.
\newblock \showarticletitle{Pythia: A Suite for Analyzing Large Language Models Across Training and Scaling}.
\newblock


\bibitem[Brown et~al\mbox{.}(2020)]%
        {gpt3}
\bibfield{author}{\bibinfo{person}{Tom~B. Brown}, \bibinfo{person}{Benjamin Mann}, \bibinfo{person}{Nick Ryder}, \bibinfo{person}{Melanie Subbiah}, \bibinfo{person}{Jared Kaplan}, \bibinfo{person}{Prafulla Dhariwal}, \bibinfo{person}{Arvind Neelakantan}, \bibinfo{person}{Pranav Shyam}, \bibinfo{person}{Girish Sastry}, \bibinfo{person}{Amanda Askell}, \bibinfo{person}{Sandhini Agarwal}, \bibinfo{person}{Ariel Herbert-Voss}, \bibinfo{person}{Gretchen Krueger}, \bibinfo{person}{T.~J. Henighan}, \bibinfo{person}{Rewon Child}, \bibinfo{person}{Aditya Ramesh}, \bibinfo{person}{Daniel~M. Ziegler}, \bibinfo{person}{Jeff Wu}, \bibinfo{person}{Clemens Winter}, \bibinfo{person}{Christopher Hesse}, \bibinfo{person}{Mark Chen}, \bibinfo{person}{Eric Sigler}, \bibinfo{person}{Mateusz Litwin}, \bibinfo{person}{Scott Gray}, \bibinfo{person}{Benjamin Chess}, \bibinfo{person}{Jack Clark}, \bibinfo{person}{Christopher Berner}, \bibinfo{person}{Sam McCandlish}, \bibinfo{person}{Alec Radford}, \bibinfo{person}{Ilya Sutskever},
  {and} \bibinfo{person}{Dario Amodei}.} \bibinfo{year}{2020}\natexlab{}.
\newblock \showarticletitle{Language Models are Few-Shot Learners}.
\newblock \bibinfo{journal}{\emph{ArXiv}}  \bibinfo{volume}{abs/2005.14165} (\bibinfo{year}{2020}).
\newblock


\bibitem[Child et~al\mbox{.}(2019)]%
        {sparseformer}
\bibfield{author}{\bibinfo{person}{Rewon Child}, \bibinfo{person}{Scott Gray}, \bibinfo{person}{Alec Radford}, {and} \bibinfo{person}{Ilya Sutskever}.} \bibinfo{year}{2019}\natexlab{}.
\newblock \showarticletitle{Generating Long Sequences with Sparse Transformers}.
\newblock \bibinfo{journal}{\emph{ArXiv}}  \bibinfo{volume}{abs/1904.10509} (\bibinfo{year}{2019}).
\newblock


\bibitem[Crankshaw et~al\mbox{.}(2017)]%
        {clipper}
\bibfield{author}{\bibinfo{person}{Daniel Crankshaw}, \bibinfo{person}{Xin Wang}, \bibinfo{person}{Guilio Zhou}, \bibinfo{person}{Michael~J Franklin}, \bibinfo{person}{Joseph~E Gonzalez}, {and} \bibinfo{person}{Ion Stoica}.} \bibinfo{year}{2017}\natexlab{}.
\newblock \showarticletitle{Clipper: A Low-Latency Online Prediction Serving System}. In \bibinfo{booktitle}{\emph{14th USENIX Symposium on Networked Systems Design and Implementation (NSDI 17)}}. \bibinfo{pages}{613--627}.
\newblock


\bibitem[Dao(2023)]%
        {flashattention2}
\bibfield{author}{\bibinfo{person}{Tri Dao}.} \bibinfo{year}{2023}\natexlab{}.
\newblock \showarticletitle{FlashAttention-2: Faster Attention with Better Parallelism and Work Partitioning}.
\newblock \bibinfo{journal}{\emph{ArXiv}}  \bibinfo{volume}{abs/2307.08691} (\bibinfo{year}{2023}).
\newblock


\bibitem[Dao et~al\mbox{.}(2022)]%
        {flashattention}
\bibfield{author}{\bibinfo{person}{Tri Dao}, \bibinfo{person}{Dan Fu}, \bibinfo{person}{Stefano Ermon}, \bibinfo{person}{Atri Rudra}, {and} \bibinfo{person}{Christopher R{\'e}}.} \bibinfo{year}{2022}\natexlab{}.
\newblock \showarticletitle{Flashattention: Fast and memory-efficient exact attention with io-awareness}.
\newblock \bibinfo{journal}{\emph{Advances in Neural Information Processing Systems}}  \bibinfo{volume}{35} (\bibinfo{year}{2022}), \bibinfo{pages}{16344--16359}.
\newblock


\bibitem[Devlin et~al\mbox{.}(2019)]%
        {bert}
\bibfield{author}{\bibinfo{person}{Jacob Devlin}, \bibinfo{person}{Ming-Wei Chang}, \bibinfo{person}{Kenton Lee}, {and} \bibinfo{person}{Kristina Toutanova}.} \bibinfo{year}{2019}\natexlab{}.
\newblock \showarticletitle{BERT: Pre-training of Deep Bidirectional Transformers for Language Understanding}.
\newblock \bibinfo{journal}{\emph{ArXiv}}  \bibinfo{volume}{abs/1810.04805} (\bibinfo{year}{2019}).
\newblock


\bibitem[Frantar et~al\mbox{.}(2023)]%
        {GPTQ}
\bibfield{author}{\bibinfo{person}{Elias Frantar}, \bibinfo{person}{Saleh Ashkboos}, \bibinfo{person}{Torsten Hoefler}, {and} \bibinfo{person}{Dan Alistarh}.} \bibinfo{year}{2023}\natexlab{}.
\newblock \showarticletitle{GPTQ: Accurate Post-Training Quantization for Generative Pre-trained Transformers}.
\newblock \bibinfo{journal}{\emph{International Conference on Learning Representations (ICLR)}} (\bibinfo{year}{2023}).
\newblock


\bibitem[Gujarati et~al\mbox{.}(2020)]%
        {clockwork}
\bibfield{author}{\bibinfo{person}{Arpan Gujarati}, \bibinfo{person}{Reza Karimi}, \bibinfo{person}{Safya Alzayat}, \bibinfo{person}{Antoine Kaufmann}, \bibinfo{person}{Ymir Vigfusson}, {and} \bibinfo{person}{Jonathan Mace}.} \bibinfo{year}{2020}\natexlab{}.
\newblock \showarticletitle{Serving DNNs like Clockwork: Performance Predictability from the Bottom Up}. In \bibinfo{booktitle}{\emph{USENIX Symposium on Operating Systems Design and Implementation}}.
\newblock


\bibitem[Herlihy et~al\mbox{.}(2019)]%
        {pq}
\bibfield{author}{\bibinfo{person}{Maurice Herlihy}, \bibinfo{person}{Nir Shavit}, \bibinfo{person}{Victor Luchangco}, {and} \bibinfo{person}{Michael Spear}.} \bibinfo{year}{2019}\natexlab{}.
\newblock \showarticletitle{Priority queues}.
\newblock \bibinfo{journal}{\emph{The Art of Multiprocessor Programming}} (\bibinfo{year}{2019}).
\newblock


\bibitem[Jin et~al\mbox{.}(2023)]%
        {s3}
\bibfield{author}{\bibinfo{person}{Yunho Jin}, \bibinfo{person}{Chun-Feng Wu}, \bibinfo{person}{David Brooks}, {and} \bibinfo{person}{Gu-Yeon Wei}.} \bibinfo{year}{2023}\natexlab{}.
\newblock \showarticletitle{S3: Increasing GPU Utilization during Generative Inference for Higher Throughput}.
\newblock \bibinfo{journal}{\emph{Advances in Neural Information Processing Systems}}  \bibinfo{volume}{36} (\bibinfo{year}{2023}).
\newblock


\bibitem[Kwon et~al\mbox{.}(2023)]%
        {vLLM}
\bibfield{author}{\bibinfo{person}{Woosuk Kwon}, \bibinfo{person}{Zhuohan Li}, \bibinfo{person}{Siyuan Zhuang}, \bibinfo{person}{Ying Sheng}, \bibinfo{person}{Lianmin Zheng}, \bibinfo{person}{Cody~Hao Yu}, \bibinfo{person}{Joseph~E. Gonzalez}, \bibinfo{person}{Haotong Zhang}, {and} \bibinfo{person}{I. Stoica}.} \bibinfo{year}{2023}\natexlab{}.
\newblock \showarticletitle{Efficient Memory Management for Large Language Model Serving with PagedAttention}.
\newblock \bibinfo{journal}{\emph{Proceedings of the ACM SIGOPS 29th Symposium on Operating Systems Principles}} (\bibinfo{year}{2023}), \bibinfo{pages}{611--626}.
\newblock


\bibitem[Lin et~al\mbox{.}(2023)]%
        {AWQ}
\bibfield{author}{\bibinfo{person}{Ji Lin}, \bibinfo{person}{Jiaming Tang}, \bibinfo{person}{Haotian Tang}, \bibinfo{person}{Shang Yang}, \bibinfo{person}{Xingyu Dang}, {and} \bibinfo{person}{Song Han}.} \bibinfo{year}{2023}\natexlab{}.
\newblock \showarticletitle{AWQ: Activation-aware Weight Quantization for LLM Compression and Acceleration}.
\newblock \bibinfo{journal}{\emph{ArXiv}}  \bibinfo{volume}{abs/2306.00978} (\bibinfo{year}{2023}).
\newblock


\bibitem[Mikolov et~al\mbox{.}(2013)]%
        {word2vec}
\bibfield{author}{\bibinfo{person}{Tomas Mikolov}, \bibinfo{person}{Kai Chen}, \bibinfo{person}{Gregory~S. Corrado}, {and} \bibinfo{person}{Jeffrey Dean}.} \bibinfo{year}{2013}\natexlab{}.
\newblock \showarticletitle{Efficient Estimation of Word Representations in Vector Space}. In \bibinfo{booktitle}{\emph{International Conference on Learning Representations}}.
\newblock


\bibitem[NVIDIA(2019)]%
        {fastertransformer}
\bibfield{author}{\bibinfo{person}{NVIDIA}.} \bibinfo{year}{2019}\natexlab{}.
\newblock \bibinfo{title}{FasterTransformer}.
\newblock
\newblock
\urldef\tempurl%
\url{https://github.com/NVIDIA/FasterTransformer}
\showURL{%
\tempurl}


\bibitem[OpenAI(2022)]%
        {chatgpt}
\bibfield{author}{\bibinfo{person}{OpenAI}.} \bibinfo{year}{2022}\natexlab{}.
\newblock \bibinfo{title}{Introducting ChatGPT}.
\newblock
\newblock
\urldef\tempurl%
\url{https://openai.com/blog/chatgpt}
\showURL{%
\tempurl}


\bibitem[Ott et~al\mbox{.}(2019)]%
        {fairseq}
\bibfield{author}{\bibinfo{person}{Myle Ott}, \bibinfo{person}{Sergey Edunov}, \bibinfo{person}{Alexei Baevski}, \bibinfo{person}{Angela Fan}, \bibinfo{person}{Sam Gross}, \bibinfo{person}{Nathan Ng}, \bibinfo{person}{David Grangier}, {and} \bibinfo{person}{Michael Auli}.} \bibinfo{year}{2019}\natexlab{}.
\newblock \showarticletitle{fairseq: A Fast, Extensible Toolkit for Sequence Modeling}. In \bibinfo{booktitle}{\emph{North American Chapter of the Association for Computational Linguistics}}. \bibinfo{pages}{6151--6162}.
\newblock


\bibitem[Ouyang et~al\mbox{.}(2022)]%
        {it2}
\bibfield{author}{\bibinfo{person}{Long Ouyang}, \bibinfo{person}{Jeff Wu}, \bibinfo{person}{Xu Jiang}, \bibinfo{person}{Diogo Almeida}, \bibinfo{person}{Carroll~L. Wainwright}, \bibinfo{person}{Pamela Mishkin}, \bibinfo{person}{Chong Zhang}, \bibinfo{person}{Sandhini Agarwal}, \bibinfo{person}{Katarina Slama}, \bibinfo{person}{Alex Ray}, \bibinfo{person}{John Schulman}, \bibinfo{person}{Jacob Hilton}, \bibinfo{person}{Fraser Kelton}, \bibinfo{person}{Luke~E. Miller}, \bibinfo{person}{Maddie Simens}, \bibinfo{person}{Amanda Askell}, \bibinfo{person}{Peter Welinder}, \bibinfo{person}{Paul~Francis Christiano}, \bibinfo{person}{Jan Leike}, {and} \bibinfo{person}{Ryan~J. Lowe}.} \bibinfo{year}{2022}\natexlab{}.
\newblock \showarticletitle{Training language models to follow instructions with human feedback}.
\newblock \bibinfo{journal}{\emph{ArXiv}}  \bibinfo{volume}{abs/2203.02155} (\bibinfo{year}{2022}).
\newblock


\bibitem[Paszke et~al\mbox{.}(2019)]%
        {pytorch}
\bibfield{author}{\bibinfo{person}{Adam Paszke}, \bibinfo{person}{Sam Gross}, \bibinfo{person}{Francisco Massa}, \bibinfo{person}{Adam Lerer}, \bibinfo{person}{James Bradbury}, \bibinfo{person}{Gregory Chanan}, \bibinfo{person}{Trevor Killeen}, \bibinfo{person}{Zeming Lin}, \bibinfo{person}{Natalia Gimelshein}, \bibinfo{person}{Luca Antiga}, {et~al\mbox{.}}} \bibinfo{year}{2019}\natexlab{}.
\newblock \showarticletitle{Pytorch: An imperative style, high-performance deep learning library}.
\newblock \bibinfo{journal}{\emph{Advances in neural information processing systems}}  \bibinfo{volume}{32} (\bibinfo{year}{2019}).
\newblock


\bibitem[Pope et~al\mbox{.}(2023)]%
        {efficientscaling}
\bibfield{author}{\bibinfo{person}{Reiner Pope}, \bibinfo{person}{Sholto Douglas}, \bibinfo{person}{Aakanksha Chowdhery}, \bibinfo{person}{Jacob Devlin}, \bibinfo{person}{James Bradbury}, \bibinfo{person}{Anselm Levskaya}, \bibinfo{person}{Jonathan Heek}, \bibinfo{person}{Kefan Xiao}, \bibinfo{person}{Shivani Agrawal}, {and} \bibinfo{person}{Jeff Dean}.} \bibinfo{year}{2023}\natexlab{}.
\newblock \showarticletitle{Efficiently scaling transformer inference}.
\newblock \bibinfo{journal}{\emph{Proceedings of Machine Learning and Systems}}  \bibinfo{volume}{5} (\bibinfo{year}{2023}).
\newblock


\bibitem[Qiu et~al\mbox{.}(2024)]%
        {proxy}
\bibfield{author}{\bibinfo{person}{Haoran Qiu}, \bibinfo{person}{Weichao Mao}, \bibinfo{person}{Archit Patke}, \bibinfo{person}{Shengkun Cui}, \bibinfo{person}{Saurabh Jha}, \bibinfo{person}{Chen Wang}, \bibinfo{person}{Hubertus Franke}, \bibinfo{person}{Zbigniew~T Kalbarczyk}, \bibinfo{person}{Tamer Ba{\c{s}}ar}, {and} \bibinfo{person}{Ravishankar~K Iyer}.} \bibinfo{year}{2024}\natexlab{}.
\newblock \showarticletitle{Efficient Interactive LLM Serving with Proxy Model-based Sequence Length Prediction}.
\newblock \bibinfo{journal}{\emph{ArXiv}}  \bibinfo{volume}{abs/2404.08509} (\bibinfo{year}{2024}).
\newblock


\bibitem[Romero et~al\mbox{.}(2021)]%
        {infaas}
\bibfield{author}{\bibinfo{person}{Francisco Romero}, \bibinfo{person}{Qian Li}, \bibinfo{person}{Neeraja~J Yadwadkar}, {and} \bibinfo{person}{Christos Kozyrakis}.} \bibinfo{year}{2021}\natexlab{}.
\newblock \showarticletitle{INFaaS: Automated model-less inference serving}. In \bibinfo{booktitle}{\emph{2021 USENIX Annual Technical Conference (USENIX ATC 21)}}. \bibinfo{pages}{397--411}.
\newblock


\bibitem[Sanh et~al\mbox{.}(2019)]%
        {distillbert}
\bibfield{author}{\bibinfo{person}{Victor Sanh}, \bibinfo{person}{Lysandre Debut}, \bibinfo{person}{Julien Chaumond}, {and} \bibinfo{person}{Thomas Wolf}.} \bibinfo{year}{2019}\natexlab{}.
\newblock \showarticletitle{DistilBERT, a distilled version of BERT: smaller, faster, cheaper and lighter}.
\newblock \bibinfo{journal}{\emph{ArXiv}}  \bibinfo{volume}{abs/1910.01108} (\bibinfo{year}{2019}).
\newblock


\bibitem[Sheng et~al\mbox{.}(2023)]%
        {flexgen}
\bibfield{author}{\bibinfo{person}{Ying Sheng}, \bibinfo{person}{Lianmin Zheng}, \bibinfo{person}{Binhang Yuan}, \bibinfo{person}{Zhuohan Li}, \bibinfo{person}{Max Ryabinin}, \bibinfo{person}{Daniel~Y. Fu}, \bibinfo{person}{Zhiqiang Xie}, \bibinfo{person}{Beidi Chen}, \bibinfo{person}{Clark~W. Barrett}, \bibinfo{person}{Joseph Gonzalez}, \bibinfo{person}{Percy Liang}, \bibinfo{person}{Christopher R{\'e}}, \bibinfo{person}{Ioan~Cristian Stoica}, {and} \bibinfo{person}{Ce Zhang}.} \bibinfo{year}{2023}\natexlab{}.
\newblock \showarticletitle{High-throughput Generative Inference of Large Language Models with a Single GPU}. In \bibinfo{booktitle}{\emph{International Conference on Machine Learning}}. PMLR, \bibinfo{pages}{31094–--31116}.
\newblock


\bibitem[Silberschatz et~al\mbox{.}(2004)]%
        {va}
\bibfield{author}{\bibinfo{person}{Abraham Silberschatz}, \bibinfo{person}{Greg Gagne}, {and} \bibinfo{person}{Peter Galvin}.} \bibinfo{year}{2004}\natexlab{}.
\newblock \showarticletitle{Operating System Principles}.
\newblock


\bibitem[Taori et~al\mbox{.}(2023)]%
        {alpaca}
\bibfield{author}{\bibinfo{person}{Rohan Taori}, \bibinfo{person}{Ishaan Gulrajani}, \bibinfo{person}{Tianyi Zhang}, \bibinfo{person}{Yann Dubois}, \bibinfo{person}{Xuechen Li}, \bibinfo{person}{Carlos Guestrin}, \bibinfo{person}{Percy Liang}, {and} \bibinfo{person}{Tatsunori~B. Hashimoto}.} \bibinfo{year}{2023}\natexlab{}.
\newblock \bibinfo{title}{Stanford Alpaca: An Instruction-following LLaMA model}.
\newblock \bibinfo{howpublished}{\url{https://github.com/tatsu-lab/stanford_alpaca}}.
\newblock


\bibitem[Team(2023)]%
        {shareGPT}
\bibfield{author}{\bibinfo{person}{ShareGPT Team}.} \bibinfo{year}{2023}\natexlab{}.
\newblock \bibinfo{title}{ShareGPT}.
\newblock
\newblock
\urldef\tempurl%
\url{https://sharegpt.com/}
\showURL{%
\tempurl}


\bibitem[Tillet et~al\mbox{.}(2019)]%
        {openaitriton}
\bibfield{author}{\bibinfo{person}{Philippe Tillet}, \bibinfo{person}{Hsiang-Tsung Kung}, {and} \bibinfo{person}{David~D. Cox}.} \bibinfo{year}{2019}\natexlab{}.
\newblock \showarticletitle{Triton: an intermediate language and compiler for tiled neural network computations}.
\newblock \bibinfo{journal}{\emph{Proceedings of the 3rd ACM SIGPLAN International Workshop on Machine Learning and Programming Languages}} (\bibinfo{year}{2019}).
\newblock


\bibitem[Touvron et~al\mbox{.}(2023)]%
        {llama2}
\bibfield{author}{\bibinfo{person}{Hugo Touvron}, \bibinfo{person}{Louis Martin}, \bibinfo{person}{Kevin~R. Stone}, \bibinfo{person}{Peter Albert}, \bibinfo{person}{Amjad Almahairi}, \bibinfo{person}{Yasmine Babaei}, \bibinfo{person}{Nikolay Bashlykov}, \bibinfo{person}{Soumya Batra}, \bibinfo{person}{Prajjwal Bhargava}, \bibinfo{person}{Shruti Bhosale}, \bibinfo{person}{Daniel~M. Bikel}, \bibinfo{person}{Lukas Blecher}, \bibinfo{person}{Cristian~Canton Ferrer}, \bibinfo{person}{Moya Chen}, \bibinfo{person}{Guillem Cucurull}, \bibinfo{person}{David Esiobu}, \bibinfo{person}{Jude Fernandes}, \bibinfo{person}{Jeremy Fu}, \bibinfo{person}{Wenyin Fu}, \bibinfo{person}{Brian Fuller}, \bibinfo{person}{Cynthia Gao}, \bibinfo{person}{Vedanuj Goswami}, \bibinfo{person}{Naman Goyal}, \bibinfo{person}{Anthony~S. Hartshorn}, \bibinfo{person}{Saghar Hosseini}, \bibinfo{person}{Rui Hou}, \bibinfo{person}{Hakan Inan}, \bibinfo{person}{Marcin Kardas}, \bibinfo{person}{Viktor Kerkez}, \bibinfo{person}{Madian Khabsa},
  \bibinfo{person}{Isabel~M. Kloumann}, \bibinfo{person}{A.~V. Korenev}, \bibinfo{person}{Punit~Singh Koura}, \bibinfo{person}{Marie-Anne Lachaux}, \bibinfo{person}{Thibaut Lavril}, \bibinfo{person}{Jenya Lee}, \bibinfo{person}{Diana Liskovich}, \bibinfo{person}{Yinghai Lu}, \bibinfo{person}{Yuning Mao}, \bibinfo{person}{Xavier Martinet}, \bibinfo{person}{Todor Mihaylov}, \bibinfo{person}{Pushkar Mishra}, \bibinfo{person}{Igor Molybog}, \bibinfo{person}{Yixin Nie}, \bibinfo{person}{Andrew Poulton}, \bibinfo{person}{Jeremy Reizenstein}, \bibinfo{person}{Rashi Rungta}, \bibinfo{person}{Kalyan Saladi}, \bibinfo{person}{Alan Schelten}, \bibinfo{person}{Ruan Silva}, \bibinfo{person}{Eric~Michael Smith}, \bibinfo{person}{R. Subramanian}, \bibinfo{person}{Xia Tan}, \bibinfo{person}{Binh Tang}, \bibinfo{person}{Ross Taylor}, \bibinfo{person}{Adina Williams}, \bibinfo{person}{Jian~Xiang Kuan}, \bibinfo{person}{Puxin Xu}, \bibinfo{person}{Zhengxu Yan}, \bibinfo{person}{Iliyan Zarov}, \bibinfo{person}{Yuchen Zhang},
  \bibinfo{person}{Angela Fan}, \bibinfo{person}{Melanie Kambadur}, \bibinfo{person}{Sharan Narang}, \bibinfo{person}{Aurelien Rodriguez}, \bibinfo{person}{Robert Stojnic}, \bibinfo{person}{Sergey Edunov}, {and} \bibinfo{person}{Thomas Scialom}.} \bibinfo{year}{2023}\natexlab{}.
\newblock \showarticletitle{Llama 2: Open Foundation and Fine-Tuned Chat Models}.
\newblock \bibinfo{journal}{\emph{ArXiv}}  \bibinfo{volume}{abs/2307.09288} (\bibinfo{year}{2023}).
\newblock


\bibitem[Vaswani et~al\mbox{.}(2017)]%
        {attention}
\bibfield{author}{\bibinfo{person}{Ashish Vaswani}, \bibinfo{person}{Noam~M. Shazeer}, \bibinfo{person}{Niki Parmar}, \bibinfo{person}{Jakob Uszkoreit}, \bibinfo{person}{Llion Jones}, \bibinfo{person}{Aidan~N. Gomez}, \bibinfo{person}{Lukasz Kaiser}, {and} \bibinfo{person}{Illia Polosukhin}.} \bibinfo{year}{2017}\natexlab{}.
\newblock \showarticletitle{Attention is All you Need}.
\newblock \bibinfo{journal}{\emph{ArXiv}}  \bibinfo{volume}{abs/1706.03762} (\bibinfo{year}{2017}).
\newblock


\bibitem[Wang et~al\mbox{.}(2023)]%
        {openchat}
\bibfield{author}{\bibinfo{person}{Guan Wang}, \bibinfo{person}{Sijie Cheng}, \bibinfo{person}{Xianyuan Zhan}, \bibinfo{person}{Xiangang Li}, \bibinfo{person}{Sen Song}, {and} \bibinfo{person}{Yang Liu}.} \bibinfo{year}{2023}\natexlab{}.
\newblock \showarticletitle{OpenChat: Advancing Open-source Language Models with Mixed-Quality Data}.
\newblock \bibinfo{journal}{\emph{ArXiv}}  \bibinfo{volume}{abs/2309.11235} (\bibinfo{year}{2023}).
\newblock


\bibitem[Wolf et~al\mbox{.}(2019)]%
        {huggingface}
\bibfield{author}{\bibinfo{person}{Thomas Wolf}, \bibinfo{person}{Lysandre Debut}, \bibinfo{person}{Victor Sanh}, \bibinfo{person}{Julien Chaumond}, \bibinfo{person}{Clement Delangue}, \bibinfo{person}{Anthony Moi}, \bibinfo{person}{Pierric Cistac}, \bibinfo{person}{Tim Rault}, \bibinfo{person}{R{\'e}mi Louf}, \bibinfo{person}{Morgan Funtowicz}, {and} \bibinfo{person}{Jamie Brew}.} \bibinfo{year}{2019}\natexlab{}.
\newblock \showarticletitle{HuggingFace's Transformers: State-of-the-art Natural Language Processing}.
\newblock \bibinfo{journal}{\emph{ArXiv}}  \bibinfo{volume}{abs/1910.03771} (\bibinfo{year}{2019}).
\newblock


\bibitem[Wu et~al\mbox{.}(2023)]%
        {fastserve}
\bibfield{author}{\bibinfo{person}{Bingyang Wu}, \bibinfo{person}{Yinmin Zhong}, \bibinfo{person}{Zili Zhang}, \bibinfo{person}{Gang Huang}, \bibinfo{person}{Xuanzhe Liu}, {and} \bibinfo{person}{Xin Jin}.} \bibinfo{year}{2023}\natexlab{}.
\newblock \showarticletitle{Fast distributed inference serving for large language models}.
\newblock \bibinfo{journal}{\emph{arXiv preprint arXiv:2305.05920}} (\bibinfo{year}{2023}).
\newblock


\bibitem[Yu and Jeong(2022)]%
        {orca}
\bibfield{author}{\bibinfo{person}{Gyeong-In Yu} {and} \bibinfo{person}{Joo~Seong Jeong}.} \bibinfo{year}{2022}\natexlab{}.
\newblock \showarticletitle{Orca: A Distributed Serving System for Transformer-Based Generative Models}. In \bibinfo{booktitle}{\emph{USENIX Symposium on Operating Systems Design and Implementation}}.
\newblock


\bibitem[Zhang et~al\mbox{.}(2023)]%
        {shepherd}
\bibfield{author}{\bibinfo{person}{Hong Zhang}, \bibinfo{person}{Yupeng Tang}, \bibinfo{person}{Anurag Khandelwal}, {and} \bibinfo{person}{Ion Stoica}.} \bibinfo{year}{2023}\natexlab{}.
\newblock \showarticletitle{SHEPHERD: Serving DNNs in the Wild}. In \bibinfo{booktitle}{\emph{Symposium on Networked Systems Design and Implementation}}.
\newblock


\bibitem[Zhang et~al\mbox{.}(2022)]%
        {opt}
\bibfield{author}{\bibinfo{person}{Susan Zhang}, \bibinfo{person}{Stephen Roller}, \bibinfo{person}{Naman Goyal}, \bibinfo{person}{Mikel Artetxe}, \bibinfo{person}{Moya Chen}, \bibinfo{person}{Shuohui Chen}, \bibinfo{person}{Christopher Dewan}, \bibinfo{person}{Mona Diab}, \bibinfo{person}{Xian Li}, \bibinfo{person}{Xi~Victoria Lin}, \bibinfo{person}{Todor Mihaylov}, \bibinfo{person}{Myle Ott}, \bibinfo{person}{Sam Shleifer}, \bibinfo{person}{Kurt Shuster}, \bibinfo{person}{Daniel Simig}, \bibinfo{person}{Punit~Singh Koura}, \bibinfo{person}{Anjali Sridhar}, \bibinfo{person}{Tianlu Wang}, {and} \bibinfo{person}{Luke Zettlemoyer}.} \bibinfo{year}{2022}\natexlab{}.
\newblock \showarticletitle{OPT: Open Pre-trained Transformer Language Models}.
\newblock \bibinfo{journal}{\emph{ArXiv}}  \bibinfo{volume}{abs/2205.01068} (\bibinfo{year}{2022}).
\newblock


\bibitem[Zhao and Stankovic(1989)]%
        {fcfs}
\bibfield{author}{\bibinfo{person}{Wei Zhao} {and} \bibinfo{person}{John~A. Stankovic}.} \bibinfo{year}{1989}\natexlab{}.
\newblock \showarticletitle{Performance analysis of FCFS and improved FCFS scheduling algorithms for dynamic real-time computer systems}.
\newblock \bibinfo{journal}{\emph{[1989] Proceedings. Real-Time Systems Symposium}} (\bibinfo{year}{1989}), \bibinfo{pages}{156--165}.
\newblock


\bibitem[Zhao et~al\mbox{.}(2024)]%
        {alisa}
\bibfield{author}{\bibinfo{person}{Youpeng Zhao}, \bibinfo{person}{Di Wu}, {and} \bibinfo{person}{Jun Wang}.} \bibinfo{year}{2024}\natexlab{}.
\newblock \showarticletitle{ALISA: Accelerating Large Language Model Inference via Sparsity-Aware KV Caching}.
\newblock \bibinfo{journal}{\emph{ArXiv}}  \bibinfo{volume}{abs/2403.17312} (\bibinfo{year}{2024}).
\newblock


\bibitem[Zheng et~al\mbox{.}(2023)]%
        {it1}
\bibfield{author}{\bibinfo{person}{Zangwei Zheng}, \bibinfo{person}{Xiaozhe Ren}, \bibinfo{person}{Fuzhao Xue}, \bibinfo{person}{Yang Luo}, \bibinfo{person}{Xin Jiang}, {and} \bibinfo{person}{Yang You}.} \bibinfo{year}{2023}\natexlab{}.
\newblock \showarticletitle{Response Length Perception and Sequence Scheduling: An LLM-Empowered LLM Inference Pipeline}.
\newblock \bibinfo{journal}{\emph{ArXiv}}  \bibinfo{volume}{abs/2305.13144} (\bibinfo{year}{2023}).
\newblock


\end{thebibliography}

\end{document}